\documentclass[sn-nature]{sn-jnl}

\usepackage{graphicx}%
\usepackage{multirow}%
\usepackage{amsmath,amssymb,amsfonts}%
\usepackage{amsthm}%
\usepackage{mathrsfs}%
\usepackage{xcolor}%
\usepackage{textcomp}%
\usepackage{manyfoot}%
\usepackage{booktabs}%
\usepackage{algorithm}%
\usepackage{algorithmicx}%
\usepackage{algpseudocode}%
\usepackage{listings}%
\usepackage{threeparttable}%
\usepackage{makecell}
\usepackage{bm}

\newcommand{\modelname}{TransTS}

\theoremstyle{thmstyleone}%
\theoremstyle{thmstyletwo}%

\theoremstyle{thmstylethree}%

\begin{document}

\title[Reaction-Transformation-Aware Flow Matching]{Reaction-Transformation-Aware Flow Matching for Generalizable Transition State Generation}


\author[1,2]{\fnm{Kaipeng} \sur{Zeng}}
\author[5,6,2]{\fnm{Wenxi} \sur{Zhai}}
\author[5,6,2]{\fnm{Shengrui} \sur{Xu}}
\author[3,2]{\fnm{Jie} \sur{Zhao}}
\author[3,2]{\fnm{Bowen} \sur{Li}}
\author[4,2]{\fnm{Shiyue} \sur{Wang}}

\author*[1,2]{\fnm{Junchi} \sur{Yan}}\email{yanjunchi@sjtu.edu.cn}
\author*[3,2]{\fnm{Tong} \sur{Zhu}}\email{tzhu@lps.ecnu.edu.cn}

\affil*[1]{\orgdiv{School of Computer Science \& School of Artificial Intelligence}, \orgname{Shanghai Jiao Tong University}, \orgaddress{\city{Shanghai}, \postcode{200240}, \state{Shanghai}, \country{China}}}

\affil*[2]{\orgname{Shanghai Innovation Institute}, \orgaddress{\city{Shanghai}, \postcode{200231}, \state{Shanghai}, \country{China}}}

\affil*[3]{\orgdiv{School of Chemistry and Molecular Engineering}, \orgname{East China Normal University}, \orgaddress{\city{Shanghai}, \postcode{200262}, \state{Shanghai}, \country{China}}}

\affil[4]{\orgdiv{School of Mathematical Sciences, Key Laboratory of MEA, and Shanghai Key Laboratory of PMMP}, \orgname{East China Normal University}, \orgaddress{\city{Shanghai}, \postcode{200241}, \state{Shanghai}, \country{China}}}

\affil[5] {\orgdiv{School of Artificial Intelligence and Data Science}, \orgname{University of Science and Technology of China}, \orgaddress{\city{Hefei}, \postcode{230026}, \state{Anhui}, \country{China}}}
\affil[6] {\orgdiv{Suzhou Institute for Advanced Research}, \orgname{University of Science and Technology of China}, \orgaddress{\city{Suzhou}, \postcode{215123}, \state{Jiangsu}, \country{China}}}


\abstract{Transition-state (TS) structures define the energetic barriers and mechanistic pathways of elementary chemical reactions, yet their identification remains computationally demanding because conventional saddle-point searches require expensive quantum-mechanical calculations. Recent machine-learning approaches have accelerated TS generation by predicting structures from reaction endpoint information, but they primarily learn geometric correspondence between endpoints and TSs, leaving the structural transformations underlying elementary reactions implicitly represented.  To address this limitation, we introduce {\modelname}, a reaction-transformation-aware framework for generalizable TS generation from atom-mapped reactant--product pairs. {\modelname} explicitly learns atom-level structural transformations between reaction endpoints and integrates them with a unified atom-aligned geometric representation of reactants, TSs and products, enabling reaction-aware equivariant generation of TS geometries. {\modelname} is designed to provide reliable TS initial guesses for subsequent quantum-chemical refinement, where generated structures are evaluated not only by geometric similarity but also by their ability to converge to validated saddle points and recover the intended reaction pathways. Across IID and zero-shot OOD benchmarks, {\modelname} demonstrates improved TS initialization quality, with particularly strong generalization to unseen reaction distributions. On the challenging GDB-10-rxn and GDB-17-rxn OOD benchmarks, {\modelname} generates TS candidates that more frequently converge to validated saddle points and recover the intended elementary reactions after refinement than existing approaches under the same training regime; for example, when trained only on Transition1x, {\modelname} recovers the intended reaction in 55.1\% of GDB-10-rxn reactions, 13.3 percentage points higher than the second-best method. Scaling reaction coverage and model capacity further improves both geometric fidelity and refinement outcomes. These results establish reaction-transformation-aware generation as a promising strategy for accelerating computational exploration of chemical reaction spaces.}




\maketitle

\section{Introduction}\label{sec1}
Transition states (TSs) are first-order saddle points on the potential energy surface (PES) that define the energetic bottleneck along the minimum-energy pathway connecting reactant and product states in elementary reactions~\cite{dewar1984location}. Accurately locating a TS determines the activation barrier that governs reaction kinetics, while its geometry reveals the structural changes associated with the elementary reaction step and thereby helps elucidate reaction mechanisms~\cite{ ismail2022successes, vanden2010transition, 1996Current}. Such mechanistic and kinetic information supports applications ranging from the construction and exploration of complex reaction networks~\cite{2018Exploration, 2018Methods, xu_2026_units} to the rational design of catalysts~\cite{klucznik2024computational, 2024Exploring, 2021Computational}.

Direct experimental characterization of TSs remains challenging because these structures are highly transient and exist only for extremely short timescales. Although techniques such as ultrafast electron diffraction have enabled direct observation of transient molecular structures~\cite{liu2023rehybridization}, these remain specialized methods and do not yet provide a routine, general route for exploring diverse chemical reactions~\cite{centurion2022ultrafast}. As a result, TS identification in practice primarily relies on quantum-mechanical calculations that locate first-order saddle points on the potential energy surface. Conventional workflows combine electronic-structure methods, such as density functional theory (DFT)~\cite{2017Thirty}, with a broad range of PES exploration and saddle-point optimization algorithms. Pathway-based approaches, including the nudged elastic band method, string method, and growing string method, search for reaction pathways and approximate transition structures between relevant configurations~\cite{doi:10.1142/9789812839664_0016,weinan2002string,peters2004growing}. Alternatively, automated PES exploration approaches, such as the artificial force induced reaction method and stochastic surface walking, aim to discover possible reaction pathways without relying solely on predefined paths~\cite{2013Exploring,2013Stochastic}. Candidate TS structures are then refined using local saddle-point optimization methods, including Hessian-based approaches and modern saddle optimizers~\cite{denzel2020hessian,hermes2022sella}. However, these approaches require repeated, expensive energy and force evaluations and can suffer from convergence difficulties on complex potential energy landscapes. Consequently, practical applications that require many TS searches, such as reaction exploration involving thousands of elementary steps~\cite{van2020kinbot,von2020exploring,margraf2023exploring}, remain computationally demanding.

The emergence of machine learning (ML) has provided new routes to accelerate transition-state exploration. One direction integrates machine-learned interatomic potentials (MLIPs) with conventional search algorithms, replacing costly quantum-mechanical evaluations of the potential energy surface while retaining iterative pathway or saddle-point optimization~\cite{schreiner2022neuralneb,yuan2024analytical}. Although substantially more efficient, their reliability depends on how accurately and transferably the learned potential reproduces the relevant region of the potential energy surface~\cite{deringer2019machine,unke2021machine}. A complementary direction aims to accelerate TS identification by directly generating candidate structures from reaction information. Early studies explored deterministic TS geometry prediction, either by directly predicting TS coordinates~\cite{jackson2021tsnet} or by learning interatomic-distance representations followed by coordinate reconstruction~\cite{pattanaik2020generating,choi2023prediction}. Conditional generative models~\cite{goodfellow2020generative,pmlr-v37-sohl-dickstein15,ho2020denoising} were subsequently explored to characterize the distributions of plausible TS structures. For example, TS-GAN used adversarial learning to generate TS candidates from reaction endpoints~\cite{makos2021generative}, while OA-ReactDiff leveraged object-aware SE(3)-equivariant diffusion for reaction-level TS generation~\cite{duan2023accurate}. More recently, flow-based approaches have formulated TS generation as continuous transport processes, offering improved sampling efficiency and controllability. These approaches differ in their source states and geometric representations: MolGEN uses Gaussian priors for conditional reaction generation~\cite{tuo2026flow}, whereas React-OT and TS-DFM use endpoint-derived source states, with React-OT transporting an interpolated reactant-product geometry in Cartesian space and TS-DFM evolving an initial distance matrix in molecular distance-geometry space~\cite{duan2025optimal,luo2026generative}.

Taken together, these advances have broadened ML-based TS generation, but they still largely formulate the task as learning the geometric correspondence between reaction endpoints and TS structures, while the structural transformation associated with elementary reactions remains implicitly represented. These transformations include bond rearrangements and changes in atomic environments between atom-corresponded reactants and products. They provide chemically meaningful information about reactive regions beyond endpoint geometry alone. Recent reaction representation studies have demonstrated the value of atom-corresponded representations for capturing such transformations in reaction understanding tasks~\cite{Zeng2026-rz, pmlr-v267-jian25b}. However, the incorporation of transformation-aware representations into 3D TS generation remains unexplored. Moreover, existing geometric formulations do not generally exploit a unified atom-aligned representation of reactants, TSs, and products that enables direct cross-state exchange of higher-order equivariant information. These limitations motivate the development of reaction-aware geometric models for more reliable and generalizable TS generation.

Here, we introduce {\modelname}, a reaction-transformation-aware framework for generalizable transition-state generation from atom-mapped reactant-product pairs. {\modelname} explicitly learns structural transformations underlying elementary reactions through a reaction-aware representation module that extracts atom-wise changes between corresponding reactant and product environments and provides transformation-informed conditions for TS generation. Furthermore, by exploiting the intrinsic atom correspondence of elementary reactions, {\modelname} represents reactants, TSs and products within a unified aligned geometric framework, enabling cross-state exchange of higher-order equivariant information throughout the generation process. Beyond coordinate-level similarity, we assess whether generated structures can support recovery of the intended elementary reaction after refinement, consistent with the common use of ML-based TS generators as initial-guess providers for quantum-chemical refinement. Across IID evaluations, {\modelname} produces candidates that remain effective for downstream refinement; after RGD1 augmentation, it also achieves the best coordinate-level accuracy. Under zero-shot OOD settings, {\modelname} shows enhanced generalization to unseen reaction distributions by achieving higher rates of successful saddle-point refinement and reaction recovery than existing approaches. Scaling reaction coverage and model capacity further improves both geometric fidelity and refinement outcomes, highlighting the potential of transformation-aware TS generation for exploring broader reaction spaces.


\section{Results}\label{sec2}

\subsection{Reaction-transformation-aware transition-state generation}

\begin{figure*}[htbp]
    \centering
    \includegraphics[width=\textwidth]{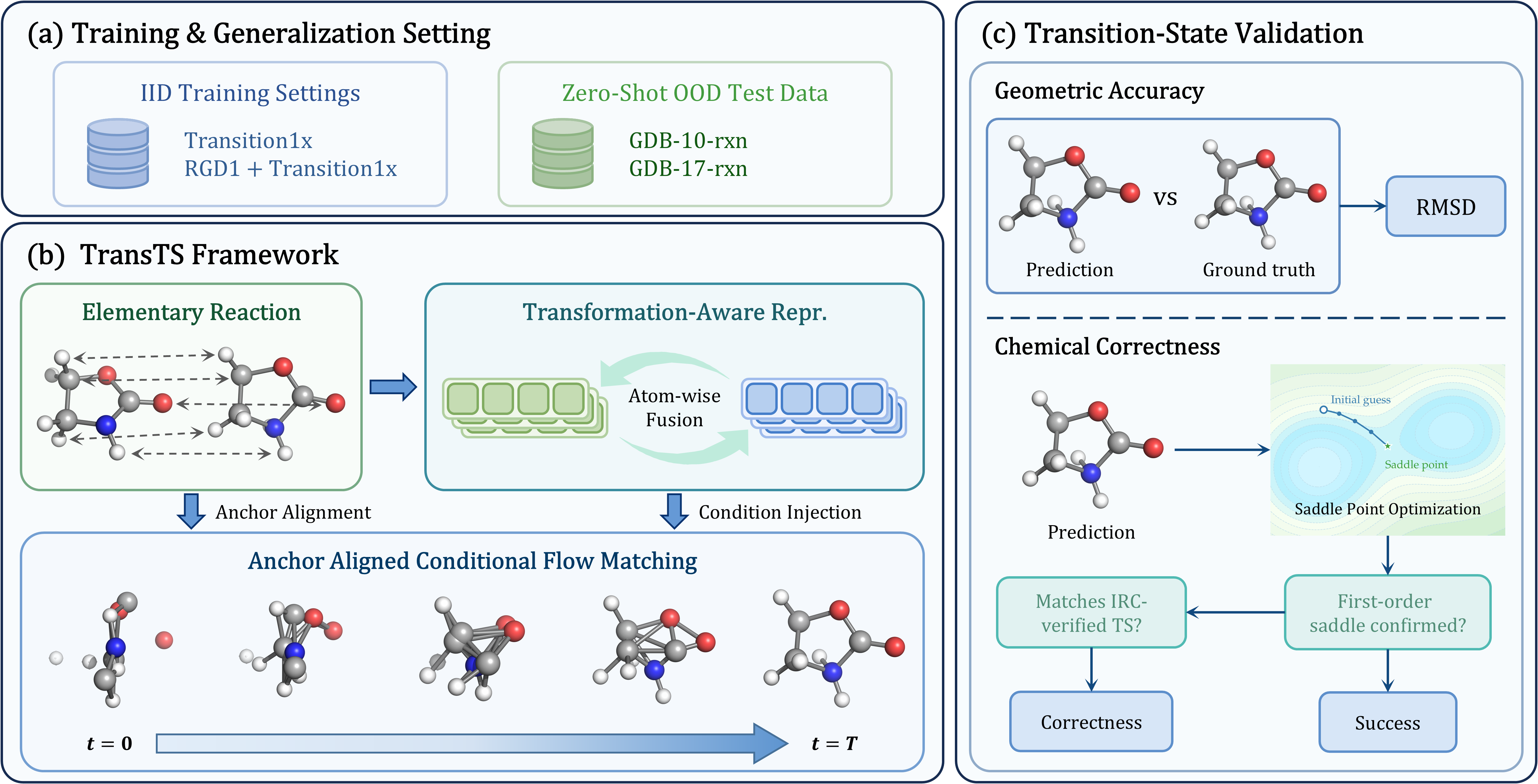}
    \caption{\textbf{Overview of \modelname{} for reaction-transformation-aware transition-state generation and validation.}
    The framework integrates reaction-transformation-aware representation learning, conditional transition-state generation and quantum-chemical validation. \textbf{a}, Training and generalization setting. \modelname{} is trained under IID settings using Transition1x or the enlarged Transition1x+RGD1 dataset, and evaluated on the corresponding Transition1x test set as well as zero-shot OOD reaction datasets, including GDB-10-rxn and GDB-17-rxn. \textbf{b}, \modelname{} framework. Atom-mapped reactant and product structures are first encoded into reaction-transformation-aware representations to capture atom-wise structural changes between reaction endpoints. The extracted reaction context is subsequently injected into an anchor-aligned conditional flow matching framework, which evolves noisy initial states into transition-state geometries. \textbf{c}, Transition-state validation. Generated structures are first evaluated by pre-optimization geometric accuracy using RMSD against reference transition states. Quantum-chemical refinement is then performed through saddle-point optimization and intrinsic reaction coordinate analysis, where optimization-success and reaction-correctness rates quantify first-order saddle-point identification and recovery of the intended elementary reaction, respectively.}
    \label{fig:overview}
\end{figure*}

We first describe the overall {\modelname} workflow for reaction-transformation-aware TS initialization and validation (Fig.~\ref{fig:overview}). The goal of TS generation is not simply to reproduce a molecular conformation, but to provide a structure that lies close enough to the intended first-order saddle point to initialize quantum-chemical refinement. We therefore designed the model around a task-specific observation: elementary reactions are characterized by specific structural rearrangements between atom-corresponded reactant and product states, and recent endpoint-based TS-generation models have shown that bond-breaking/forming information and locally important interatomic distances provide strong cues for reconstructing viable TS guesses~\cite{bickelhaupt2017analyzing,pattanaik2020generating}. A representation that treats all endpoint geometry as uniformly informative may fit the standard benchmark but can still underemphasize the structural transformation that distinguishes one elementary reaction channel from another.

\begin{figure*}[htbp]
    \centering
    \includegraphics[width=\textwidth]{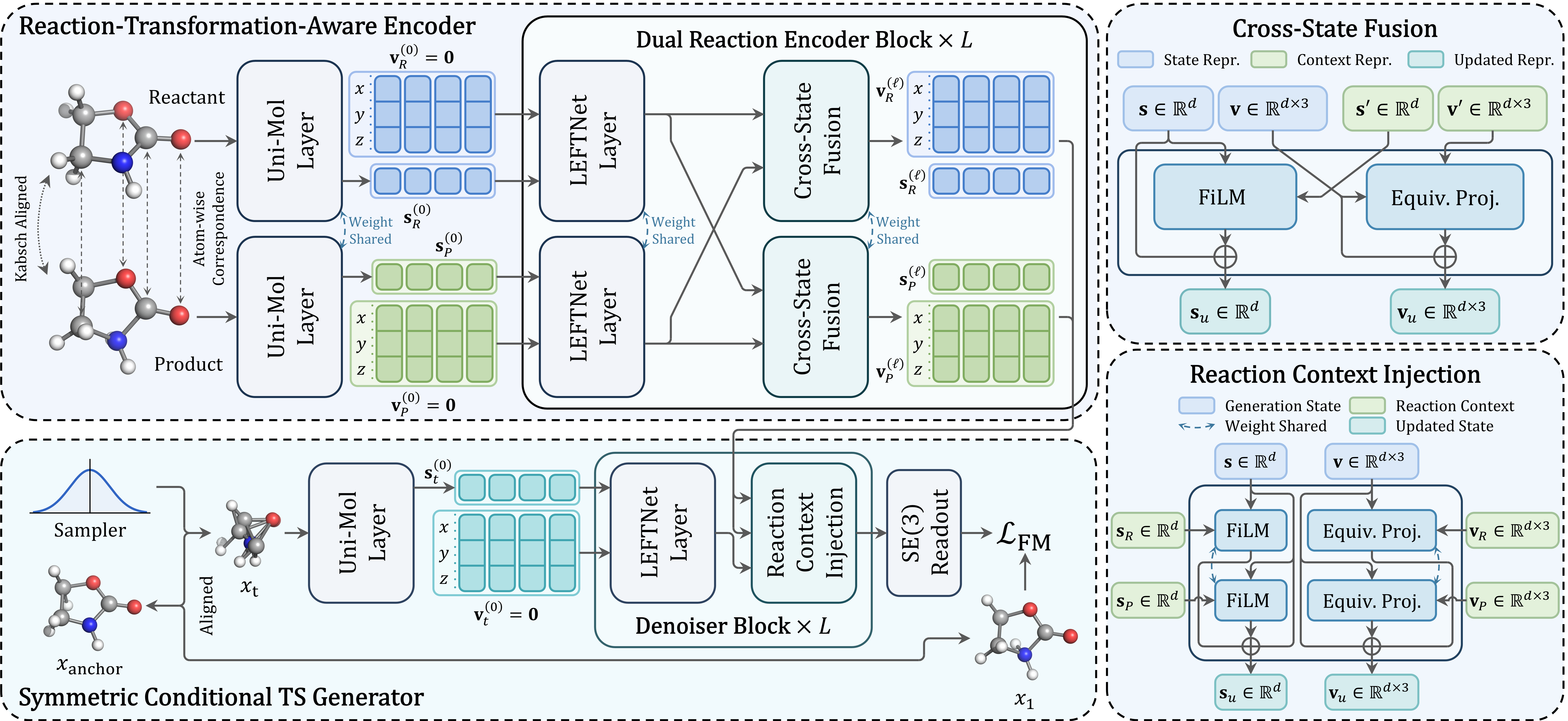}
    \caption{\textbf{Model architecture for reaction-transformation-aware TS generation.}
    The architecture is organized into a reaction-transformation-aware dual encoder, a time-conditioned equivariant TS generator, and two expanded interaction modules. In the upper-left block, atom-mapped reactant and product conformations are encoded by shared-parameter branches, with Cross-State Fusion exchanging information between corresponding atoms to make endpoint-dependent structural changes explicit. In the lower-left block, the generator evolves the noisy TS representation as a generation state and conditions it on the encoded reactant and product contexts through symmetric Reaction-Context Injection. The right-side blocks expand the two key interactions: Cross-State Fusion updates endpoint state representations using counterpart context representations, whereas Reaction-Context Injection integrates reactant and product contexts into the TS generation state through scalar FiLM modulation and equivariant vector fusion.}
    \label{fig:model_architecture}
\end{figure*}

The proposed model addresses this issue by learning a reaction-transformation-aware condition from the paired reactant and product structures. Given atom-mapped endpoints, a shared-parameter dual encoder repeatedly applies Cross-State Fusion between corresponding reactant and product atoms (Fig.~\ref{fig:model_architecture}). This module makes endpoint differences explicit to the network: atoms in conserved scaffold regions provide stable geometric context, while atoms whose local environments differ across endpoints receive reaction-specific conditioning. The resulting endpoint representations are supplied to a time-conditioned equivariant generator through symmetric reaction-context injection, where reactant and product contexts condition the evolving TS generation state. This design allows the generative dynamics to allocate modeling capacity to reaction-relevant structural changes rather than relearning the entire molecular geometry from scratch. The generator is further coupled to an endpoint-symmetric anchor-aligned flow, which reduces the source-target transport burden while preserving training-inference consistency. Full architectural and mathematical details are provided in Methods.

\subsection{Evaluation protocol for TS initial guesses}

We evaluate the proposed model under two training regimes. In the first, all models are trained only on Transition1x~\cite{schreiner2022transition1x}, the most widely used public benchmark for deep-learning-based TS generation. In the second, the training data are expanded with the larger RGD1 reaction dataset~\cite{zhao2023comprehensive}. This second regime follows recent scaling studies in TS generation~\cite{duan2025optimal}, where larger reaction corpora are introduced to test whether models can convert broader reaction diversity into improved TS initialization. To make this scaling comparison fair, all baselines are enlarged under the same parameter-growth principle rather than comparing a scaled version of the proposed model against fixed-capacity baselines.

The trained models are tested in three inference settings. We first evaluate on the independent and identically distributed (IID) Transition1x test split to maintain direct comparability with prior studies. We then perform zero-shot inference on two out-of-distribution (OOD) datasets, GDB-10-rxn and GDB-17-rxn, derived from the GDB chemical-universe databases~\cite{GDB-13,GDB-17}, without continued training or dataset-specific adaptation. These OOD tests are designed to separate benchmark interpolation from robustness to reactions outside the training distribution. Details of the OOD dataset construction are provided in Supplementary Sec.~\ref{sec:si_ood_dataset_construction}.

For Transition1x, we report RMSD against the original provided labels to maintain comparability with prior TS-generation studies~\cite{duan2023accurate,duan2025optimal}. Using the same evaluation convention, RMSD is computed as the minimum value after atom reindexing, chirality-insensitive evaluation, and Kabsch rigid alignment~\cite{kabsch1976solution,kabsch1978discussion}. Because these labels originate from NEB-derived reaction paths and are not guaranteed to be IRC-verified saddle points for the annotated reaction in every case, we additionally evaluate a post-checked clean subset whose reference TSs are validated by saddle-point optimization and IRC endpoint matching (Supplementary Sec.~\ref{sec:si_transition1x_post_check}). The OOD benchmarks use IRC-verified reference TSs, so RMSD is reported against a single reference set.

Because a deep-learning-generated TS geometry is not guaranteed to satisfy first-order saddle-point conditions, we treat each generated structure as an initial guess for quantum-chemical refinement. We therefore evaluate each prediction at two stages: RMSD is computed on the raw generated geometry before saddle-point optimization, whereas invalid, optimization-success, and reaction-correctness rates are computed after saddle-point optimization. These post-optimization metrics separate unusable generated structures, successful convergence to any first-order saddle point, and convergence to the saddle point corresponding to the intended input reaction. Full metric definitions, invalid-structure criteria, and saddle-point refinement settings are provided in Supplementary Sec.~\ref{sec:si_evaluation_metrics}. Numerical summaries for all main performance figures are provided in Supplementary Sec.~\ref{sec:si_full_benchmark_statistics}, including RMSD statistics in Supplementary Tab.~\ref{tab:si_rmsd_statistics} and post-optimization outcome rates in Supplementary Tab.~\ref{tab:si_outcome_rates}.

\subsection{IID Transition1x evaluation shows complementary geometry and refinement behavior}

\begin{figure*}[htbp]
    \centering
    \includegraphics[width=\textwidth]{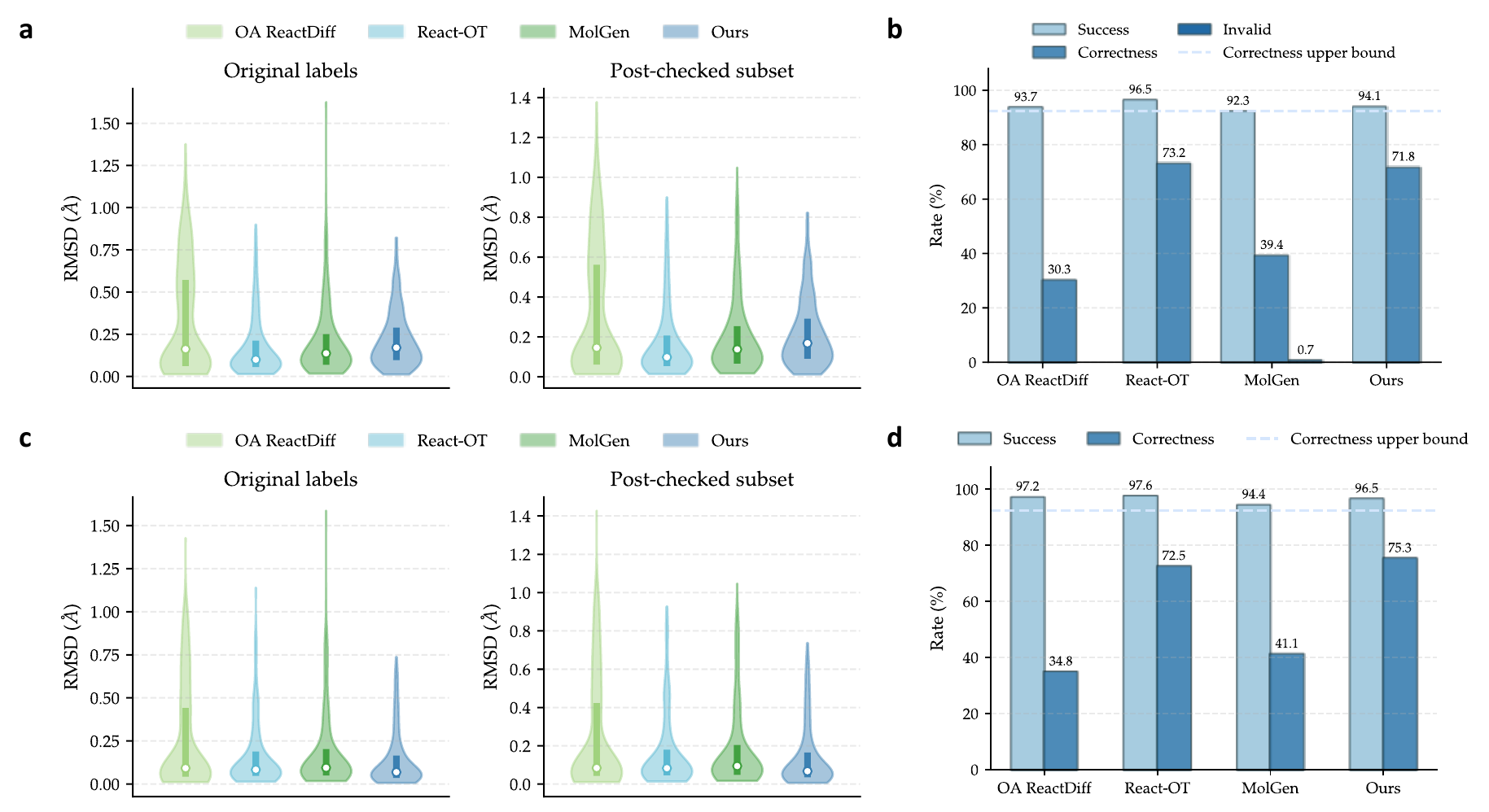}
    \caption{\textbf{Cross-model IID Transition1x performance with and without RGD1 augmentation.}
    Panels compare \modelname{} with baseline TS-generation models; T1x denotes the Transition1x dataset. \textbf{a}, Pre-optimization RMSD distributions for models trained on T1x. RMSD is reported against the original Transition1x labels and the post-checked Transition1x subset. Violin plots show per-reaction RMSD distributions; the embedded box spans Q1--Q3, where Q1 and Q3 denote the 25th and 75th percentiles, and the circle marks the median. \textbf{b}, Post-optimization outcome rates for the same T1x-trained models, including invalid, optimization-success, and reaction-correctness rates. \textbf{c}, RMSD distributions after training on T1x+RGD1, shown with the same violin, interquartile-range, and median markers as in \textbf{a}. \textbf{d}, Post-optimization outcome rates after T1x+RGD1 training, using the same outcome definitions as in \textbf{b}.}
    \label{fig:t1x_performance}
\end{figure*}

The IID Transition1x benchmark provides the standard in-distribution test for comparing TS-generation models with prior studies, with RMSD measuring pre-optimization geometry and post-optimization outcomes testing whether the generated geometry remains useful as a TS initial guess (Fig.~\ref{fig:t1x_performance}). When models are trained only on Transition1x, \modelname{} does not achieve the lowest RMSD (Fig.~\ref{fig:t1x_performance}a): its median RMSD is 0.171~\AA{} against the original labels and 0.168~\AA{} on the post-checked subset, whereas React-OT reaches 0.100~\AA{} and 0.098~\AA{}, respectively. Thus, under the smallest training regime, \modelname{} remains less competitive by this geometry-only criterion. After augmenting the training set with RGD1, however, the geometry gap is removed (Fig.~\ref{fig:t1x_performance}c): \modelname{} reaches the lowest median RMSD on both the original-label and post-checked evaluations, improving to 0.069~\AA{} and 0.067~\AA{}, respectively.

Post-optimization outcomes test a different aspect of TS-initialization quality: whether the generated geometry remains usable in saddle-point optimization and refines toward the intended reaction channel. The corresponding refinement results in Fig.~\ref{fig:t1x_performance}b show that, when trained only on Transition1x, \modelname{} remains close to the best methods in this evaluation, achieving 94.1\% optimization success and 71.8\% reaction correctness, compared with 96.5\% and 73.2\% for React-OT. With RGD1 augmentation, \modelname{} combines strong geometric accuracy with the highest reaction-correctness rate, reaching 75.3\% correctness while maintaining a 96.5\% optimization-success rate (Fig.~\ref{fig:t1x_performance}d).

These IID results separate two aspects of TS generation that are often conflated: coordinate-level agreement with a reference TS and the ability of the generated geometry to initialize refinement toward the intended reaction channel. The Transition1x-only setting shows that \modelname{} is not yet the strongest coordinate-level fitter under the smallest public-data regime, but its post-optimization behavior remains competitive, indicating that the structural-change-aware condition preserves reaction-channel information useful for practical TS initialization. With RGD1 augmentation, this reaction-aware refinement behavior is accompanied by the best IID RMSD, demonstrating that the scaled model achieves coordinate-level accuracy without losing reaction-consistent refinement.

\subsection{Zero-shot OOD evaluation shows robust TS initialization}

Generalization beyond Transition1x tests whether a TS generator has learned transferable reaction information rather than benchmark-specific geometry. We therefore evaluate zero-shot transfer to the GDB-10-rxn and GDB-17-rxn OOD benchmarks without continued training or dataset-specific adaptation. This setting is more demanding than the IID Transition1x split because the model must infer saddle-point geometries for reactions whose molecular size and molecular formula distributions shift away from the standard benchmark. GDB-10-rxn already contains larger systems than the Transition1x test set and samples a different distribution of CHNO stoichiometries, whereas GDB-17-rxn further extends the evaluation to substantially larger and more complex molecular systems. Because the OOD reference TSs are IRC-verified, RMSD is reported against a single reference set.

\begin{figure*}[htbp]
    \centering
    \includegraphics[width=\textwidth]{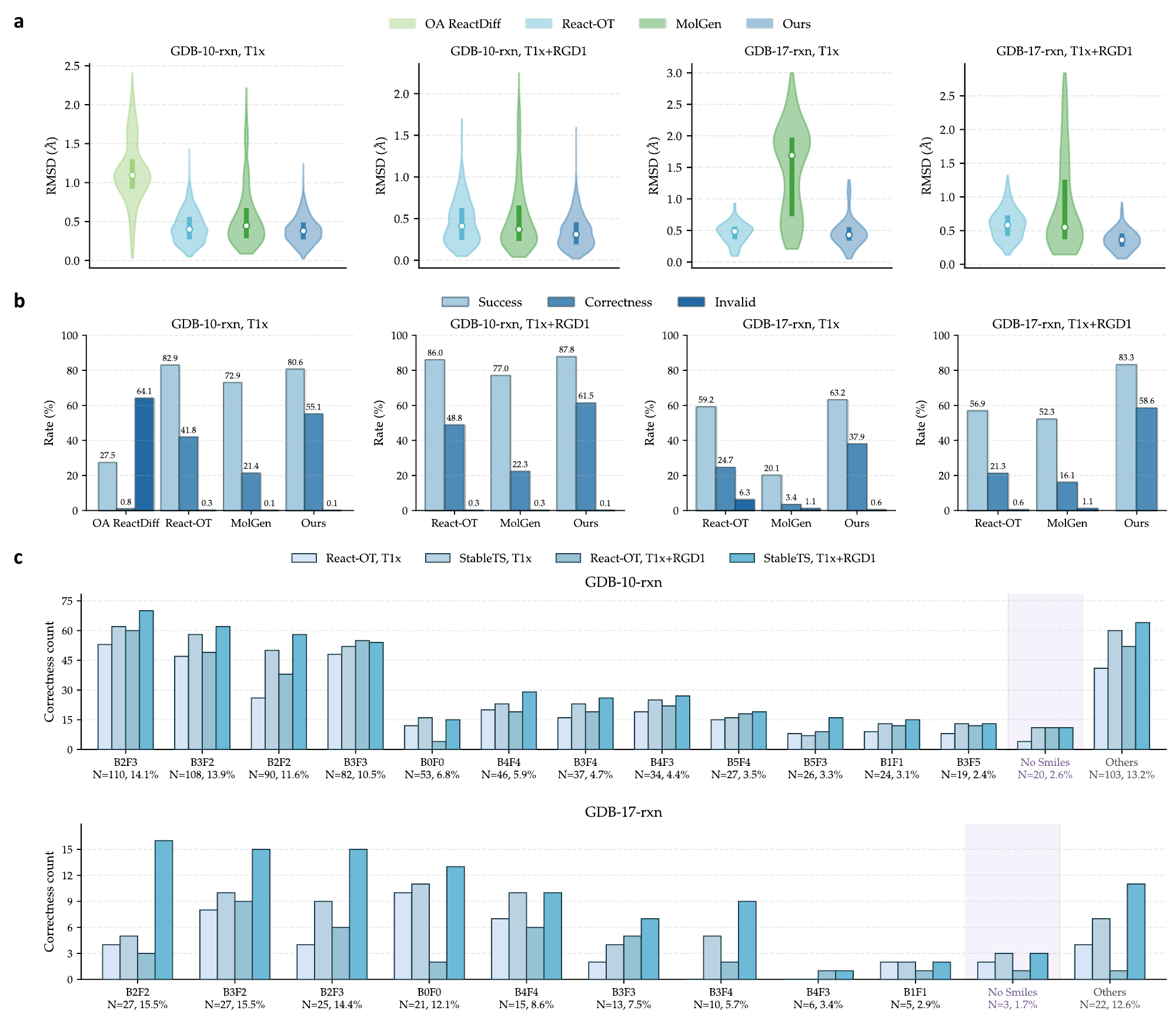}
    \caption{\textbf{Zero-shot OOD performance on GDB-10-rxn and GDB-17-rxn.}
    Panels compare \modelname{} with baseline TS-generation models under the indicated OOD test dataset and training set; T1x denotes the Transition1x dataset. OA-ReactDiff is not shown for GDB-10-rxn after T1x+RGD1 training or for either GDB-17-rxn training setting because it produced no valid structures in these settings. \textbf{a}, Pre-optimization RMSD distributions on GDB-10-rxn and GDB-17-rxn for models trained on T1x or T1x+RGD1. Violin plots show per-reaction RMSD distributions; the embedded box spans Q1--Q3, where Q1 and Q3 denote the 25th and 75th percentiles, and the circle marks the median. \textbf{b}, Post-optimization outcome rates for the same OOD settings, including invalid, optimization-success, and reaction-correctness rates. \textbf{c}, Reaction-correctness counts for \modelname{} and React-OT stratified by graph-defined B$n$F$m$ bond-rearrangement class, where $n$ and $m$ denote the numbers of broken and formed bonds, respectively. B0F0 denotes cases without graph-level bond breakage, formation, or bond-order change, but can still include conformational or stereochemical endpoint rearrangements; \texttt{NO\_SMILES} and \texttt{Other} follow the stratification protocol in Supplementary Sec.~\ref{sec:si_ood_bnfm_stratification}.}
    \label{fig:ood_performance}
\end{figure*}

When trained only on Transition1x, \modelname{} already provides the strongest OOD profile across pre-optimization geometry and post-optimization correctness (Fig.~\ref{fig:ood_performance}). Its median RMSD is the lowest on both benchmarks, reaching 0.379~\AA{} on GDB-10-rxn and 0.429~\AA{} on GDB-17-rxn, compared with the second-best values of 0.401~\AA{} and 0.485~\AA{} from React-OT. The separation becomes clearer after saddle-point optimization, particularly for reaction-correctness, which is the most direct post-refinement test of whether an initial guess follows the intended reaction channel. \modelname{} achieves 55.1\% reaction correctness on GDB-10-rxn and 37.9\% on GDB-17-rxn, exceeding the next-best model by 13.3 and 13.2 percentage points, respectively. Its optimization-success rate is also the best on GDB-17-rxn and competitive on GDB-10-rxn.

After augmenting the training set with RGD1, the OOD margins widen. \modelname{} improves the median RMSD to 0.310~\AA{} on GDB-10-rxn and 0.362~\AA{} on GDB-17-rxn, corresponding to absolute gains of 0.057~\AA{} and 0.202~\AA{} over the second-best coordinate-level baseline, MolGEN. The post-optimization outcomes follow the same trend. \modelname{} achieves the highest optimization-success and reaction correctness rates in both OOD settings, with reaction correctness increasing to 61.5\% on GDB-10-rxn and 58.6\% on GDB-17-rxn. These values exceed the next-best post-optimization baseline, React-OT, by 12.7 and 37.3 percentage points, respectively. The gain is most striking on GDB-17-rxn, where increased molecular size and structural complexity make the benchmark a stringent test of transfer beyond the standard Transition1x scale.

To further assess whether the OOD advantage is associated with reaction-transformation modeling, we stratified the OOD reactions by graph-defined B$n$F$m$ bond-rearrangement class (Fig.~\ref{fig:ood_performance}c). The stratified comparison shows that the OOD advantage is broadly distributed rather than driven by a single dominant bond-rearrangement class. \modelname{} gives higher reaction-correctness counts than React-OT in most retained B$n$F$m$ classes, often by a clear margin, and the gain persists in higher-rearrangement classes and in the long-tail \texttt{Other} group, which collects sparse bond-change patterns beyond the major classes.

Together, these OOD results show the advantage of explicitly modeling reaction-associated structural changes. Across both OOD benchmarks, \modelname{} produces closer pre-optimization TS geometries and more often refines to the intended reaction channel after saddle-point optimization. This joint improvement indicates that \modelname{} provides more reliable TS initial guesses under distribution shift, improving both pre-optimization geometry and post-refinement recovery of the intended reaction channel.

\subsection{Scaling with broader reaction coverage and model capacity}

\begin{figure*}[htbp]
    \centering
    \includegraphics[width=\textwidth]{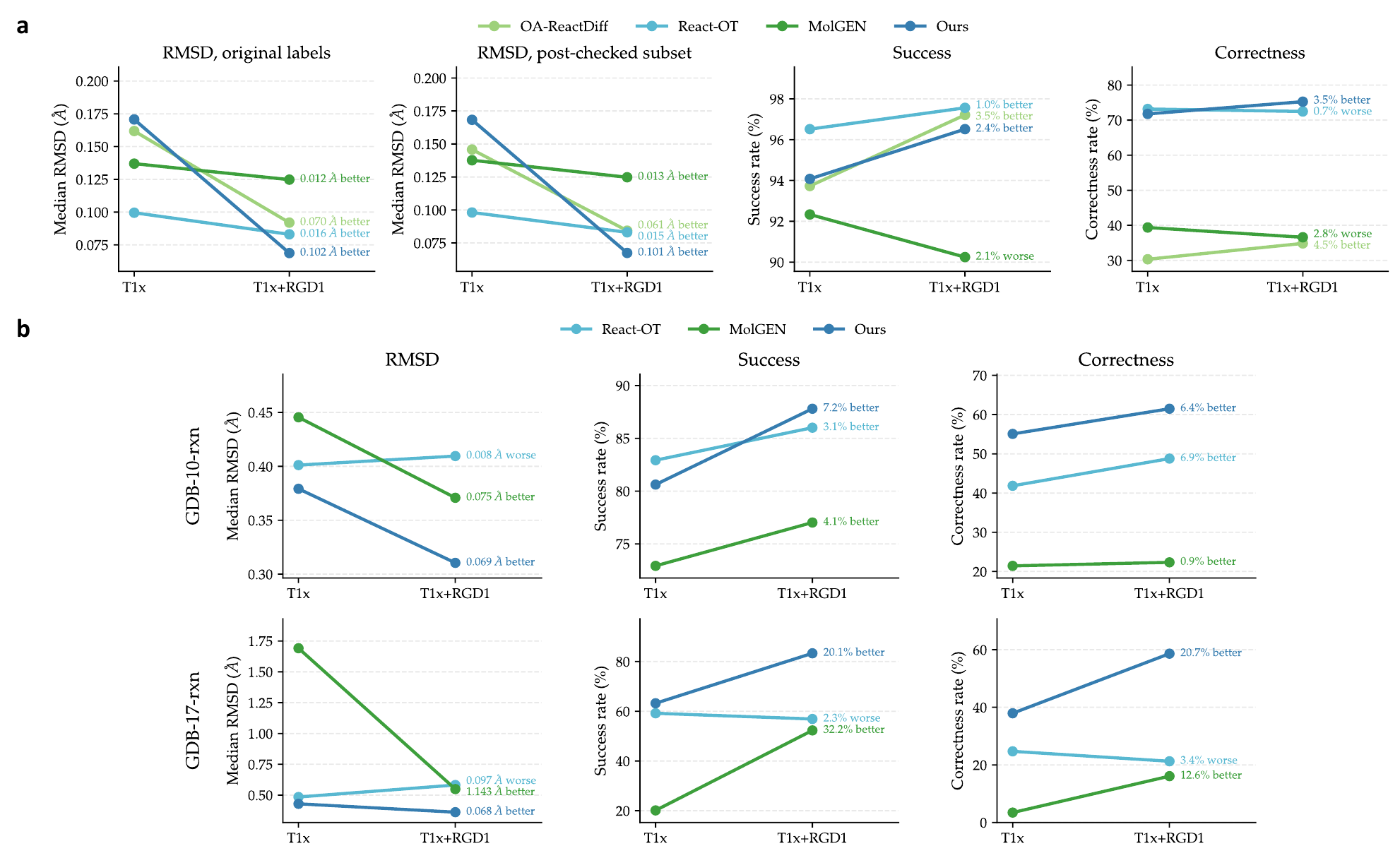}
    \caption{\textbf{Scaling behavior from Transition1x to Transition1x+RGD1 training.}
    Panels summarize the change from Transition1x-only training to Transition1x+RGD1 training. \textbf{a}, Scaling behavior on Transition1x, including median RMSD against both the original labels and the post-checked subset, together with post-optimization outcome rates. \textbf{b}, Scaling behavior on the OOD benchmarks, with rows denoting GDB-10-rxn and GDB-17-rxn and columns denoting median RMSD, optimization success and reaction correctness. Paired markers connect each model across the two training regimes, showing both the scaled performance and the corresponding change from the Transition1x-only setting.}
    \label{fig:all_improvement}
\end{figure*}

To test whether broader reaction coverage and matched model scaling translate into better TS initialization, we compare each model before and after augmenting Transition1x with RGD1 and increasing model capacity under the matched scaling rule (Fig.~\ref{fig:all_improvement}). The question is whether added data and capacity improve both the initial guess and the subsequent saddle-point refinement.

The RMSD trend is consistent with this scaling view. \modelname{} shows lower median RMSD on every reported geometry evaluation, with improvements of 0.102~\AA{} against the original Transition1x labels, 0.101~\AA{} on the post-checked Transition1x subset, 0.069~\AA{} on GDB-10-rxn, and 0.068~\AA{} on GDB-17-rxn. The same trend appears after saddle-point optimization. For \modelname{}, optimization success increases by 2.4, 7.2, and 20.1 percentage points on Transition1x, GDB-10-rxn, and GDB-17-rxn, respectively. Reaction correctness also increases across all three benchmarks, by 3.5, 6.4, and 20.7 percentage points. The largest post-optimization gains occur on GDB-17-rxn, indicating that the scaled model benefits most strongly in the most demanding OOD setting.

This scaling pattern is not simply a largest-delta comparison. Several baselines also improve in selected metrics after RGD1 augmentation, but their gains are less uniform across datasets and post-optimization outcomes. By contrast, \modelname{} improves simultaneously in pre-optimization geometry and reaction-consistent refinement across the full benchmark suite. These results indicate that the reaction-transformation-aware framework can be scaled up effectively, with broader reaction diversity and larger model capacity translating into consistent improvements across the benchmark suite.

\subsection{Representative reaction-center structural evolution in generated transition states}

\begin{figure*}[htbp]
    \centering
    \includegraphics[width=\textwidth]{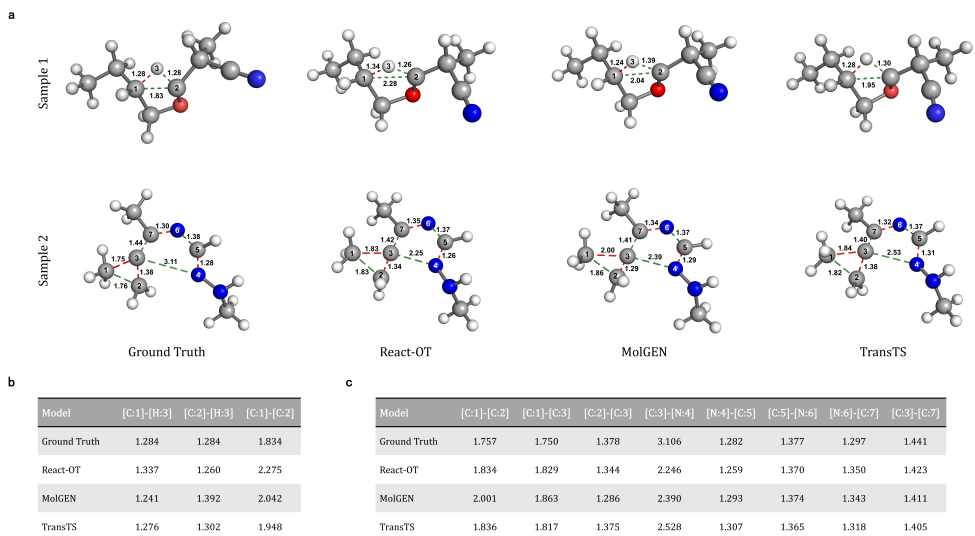}
    \caption{\textbf{Reaction-center structural evolution in zero-shot OOD transition-state generation.}
    \textbf{a}, Structural comparison for two GDB-10-rxn cases. Columns compare the reference transition state, React-OT, MolGEN and \modelname{} from left to right. Reaction-center atoms and selected interatomic distances are annotated in each structure. Red edges indicate bond cleavage or a bond-order decrease along the reaction coordinate, whereas green edges indicate bond formation or a bond-order increase. \textbf{b,c}, Summaries of selected reaction-core interatomic distances for Sample~1 and Sample~2, respectively. Distances are reported in \AA{} for atom pairs involved in the reaction-core transformation.}
    \label{fig:case_study}
\end{figure*}

To illustrate how the proposed reaction-transformation-aware representation captures local structural evolution during chemical reactions, we present two representative transition-state generation examples in Fig.~\ref{fig:case_study}. The predicted structures are generated by models trained only on Transition1x and evaluated on the zero-shot GDB-10-rxn OOD test set. Rather than emphasizing global conformational similarity, these examples examine whether the generated structures recover the spatial evolution of the key reaction-induced transformations in the transition state.

In the first example, the reaction involves the formation of the \texttt{[C:1]-[C:2]} bond, accompanied by hydrogen transfer between \texttt{[C:1]} and \texttt{[C:2]}. In the reference transition state, the \texttt{[C:1]-[C:2]} distance is 1.83~\AA{}, while the \texttt{[C:1]-[H:2]} and \texttt{[C:2]-[H:2]} distances are both 1.28~\AA{}. The nearly identical C--H distances place the hydrogen atom in a symmetric bond-breaking and bond-forming configuration, coordinated with concurrent C--C bond formation. The transition state generated by {\modelname} preserves this coupled evolution, with predicted distances of 1.95, 1.30, and 1.28~\AA{} for the three selected reaction-core distances and a reaction-core distance MAE of 0.05~\AA{}. Compared with MolGEN and React-OT, the remaining mismatch is more pronounced in the forming C--C distance, with reaction-core distance MAEs of 0.12 and 0.17~\AA{}, respectively. This example illustrates that the learned reaction condition helps preserve the geometric coordination between bond formation and hydrogen migration.

The second example represents a more complex multi-center transformation involving eight reaction-center bond or bond-order changes, including the formation of \texttt{[C:1]-[C:2]} and \texttt{[N:4]-[C:3]}, cleavage of \texttt{[C:2]-[C:3]}, and several conjugated bond-order rearrangements. In the reference transition state, the \texttt{[C:1]-[C:2]} and \texttt{[C:2]-[C:3]} distances are 1.76 and 1.75~\AA{}, respectively, indicating that the local bond-forming and bond-breaking processes are close to synchronous. By contrast, the \texttt{[N:4]-[C:3]} distance remains elongated at 3.11~\AA{}, indicating delayed bond formation and a strongly asynchronous component of the reaction transformation. The structures generated by {\modelname}, React-OT, and MolGEN achieve reaction-core distance MAEs of 0.10, 0.14, and 0.16~\AA{}, respectively. All methods recover the major reaction-center rearrangements, and the dominant uncertainty arises from the long-range \texttt{[N:4]-[C:3]} interaction, whose predicted distances are 2.53, 2.25, and 2.39~\AA{} for {\modelname}, React-OT, and MolGEN, respectively, reflecting the chemically challenging asynchronous character of this transformation. Although all methods underestimate this delayed bond formation, {\modelname} gives the closest prediction to the reference transition state. Beyond this long-range interaction, {\modelname} maintains errors below 0.09~\AA{} for the remaining selected reaction-core distances, suggesting that it better preserves the coupled local transformation pattern.

Together, these examples show that {\modelname} can recover characteristic reaction-center structural evolution, including coupled bond formation, bond cleavage, and bond-order rearrangements, while highlighting the remaining difficulty of highly asynchronous transformations involving long-range interactions.

\section{Discussion}\label{sec:discussion}

In this work, we present {\modelname}, a reaction-transformation-aware framework for transition-state (TS) generation from atom-mapped reactant and product structures. Given this mechanistic specification, {\modelname} explicitly compares corresponding atoms across the reactant and product to construct a condition that emphasizes the structural transformation associated with the elementary reaction. This reaction condition is coupled with an endpoint-symmetric canonical anchor for aligning the target TS and the sampled prior, reducing the source-target transport burden while preserving consistency between training and inference.

Across the benchmark suite, this design provides its clearest advantage under distribution shift. Although {\modelname} trained only on Transition1x is not the strongest coordinate-level fitter on the IID benchmark, it remains competitive after saddle-point refinement; after RGD1 augmentation, it also achieves the best IID RMSD. More importantly, {\modelname} demonstrates the strongest zero-shot OOD performance on both GDB-10-rxn and GDB-17-rxn, and its gains from RGD1 scaling are consistent across geometry and refinement metrics. These results are consistent with endpoint structural-change modeling helping preserve reaction-channel information under distribution shift, as generated TS guesses more often refine toward the intended reaction channel beyond the standard Transition1x distribution. This is important for practical reaction exploration, where the target chemistry is often outside the narrow distribution of public TS-generation benchmarks.

\textbf{Limitations.} Accurate coordinate-level TS generation appears to depend on data scale: when trained only on Transition1x, {\modelname} is not the strongest RMSD model; however, its refinement behavior remains competitive. Broader reaction coverage and increased capacity improve this geometric accuracy, suggesting that larger-scale training is important when high-fidelity TS coordinates are required. The current evaluation focuses on organic CHNO reaction space, and broader chemistry involving charged species, radicals, transition metals, or explicit solvent/catalyst environments will require additional data and validation.

\section{Methods}
\label{sec:methods}

\subsection{Elementary reactions as atom-aligned geometric systems}
\label{sec:prelim}

We formulate transition state (TS) generation as a conditional 3D structure generation problem. We consider atom-mapped elementary reactions characterized by a shared atom index set $\mathcal{V}=\{1,\dots,N\}$ and atomic numbers $\mathbf{a}=(a_i)_{i=1}^N$. The spatial configurations of the reactant, product, and transition state are denoted by Cartesian coordinates $\mathbf{x}^R, \mathbf{x}^P, \mathbf{x}^{TS}\in\mathbb{R}^{N\times 3}$, respectively. The objective is to learn a conditional distribution that generates the TS geometry, given the paired atom-mapped reactant and product structures.

Atom mapping specifies the elementary reaction channel: the same molecular endpoints can correspond to different mechanisms under different atom correspondences~\cite{doi:10.1021/ci3002217}. This correspondence allows the reactant, product and TS conformations to be represented as states of one reaction-level geometric system. Object-aware formulations preserve $SE(3)$ equivariance by assigning separate transformation freedoms to molecular roles such as reactant, product and transition state~\cite{duan2023accurate,tuo2026flow}. In the atom-aligned setting considered here, the elementary reaction can instead be represented in a shared frame that is equivariant to a single global rigid motion. We first map the coordinates to a center-of-mass-free (com-free) subspace:
\begin{equation*}
\bar{\mathbf{x}}^S = \mathbf{x}^S - \frac{1}{N}\mathbf{1}\mathbf{1}^\top \mathbf{x}^S,
\quad S \in \{R, P, TS\}
\end{equation*}
Next, we construct a shared canonical frame by rigidly aligning the product with the reactant using the Kabsch algorithm~\cite{kabsch1976solution, kabsch1978discussion}:
\begin{equation*}
\hat{\mathbf{x}}^P = \mathcal{A}(\bar{\mathbf{x}}^P, \bar{\mathbf{x}}^R)
\end{equation*}
where $\mathcal{A}(\mathbf{x}, \mathbf{y})$ denotes the optimal rotation that minimizes the Frobenius norm between the atom-mapped conformations.

To provide a deterministic, structure-aware reference for the generative process, we define a generalized, inferable canonical anchor within the com-free subspace:
\begin{equation*}
\mathbf{x}_{\text{anchor}} = \mathcal{F}_{\text{anchor}}(\bar{\mathbf{x}}^R, \hat{\mathbf{x}}^P)
\end{equation*}
Because reversing the reactant and product endpoints describes the same transition-state geometry, the canonical anchor should not depend on an arbitrary forward or reverse reaction direction. We therefore choose the anchor function to be symmetric with respect to the two endpoints:
\begin{equation*}
\mathcal{F}_{\text{anchor}}(\bar{\mathbf{x}}^R, \hat{\mathbf{x}}^P)
=
\mathcal{F}_{\text{anchor}}(\hat{\mathbf{x}}^P, \bar{\mathbf{x}}^R).
\end{equation*}
Thus, the canonical reference frame is aligned with the intrinsic endpoint symmetry of transition-state generation.
By subsequently aligning the transition state to this inferred anchor via the Kabsch algorithm, we establish a unique, deterministic TS conformation for any given aligned reactant-product pair:
\begin{equation*}
    \hat{\mathbf{x}}^{TS} = \mathcal{A}(\bar{\mathbf{x}}^{TS}, \mathbf{x}_{\text{anchor}})
\end{equation*}
This completes the unified reaction frame used by the representation learner, generator and flow-matching objective below.

\subsection{Molecular-state feature initialization}
\label{sec:method_init}

Before introducing the branch-specific encoder and generator, we define a common molecular-state initialization. The reactant, product and TS interpolant are each represented with scalar node features, invariant edge features and equivariant vector features. We use the same initialization design for all states before their subsequent representation-learning or generation updates.

For each atom $i$, we initialize the scalar node feature with an element embedding $\mathbf{h}_i = \mathrm{Emb}(a_i)$. For each atom pair $(i,j)$ within a spatial cutoff radius $r_c$, we compute the Euclidean distance $d_{ij}=\|\mathbf{x}_i-\mathbf{x}_j\|_2$ and construct a set of radial basis functions~\cite{duan2023accurate}:
\[
\phi_k(d_{ij}) = \frac{1}{2}\Bigl(\cos\bigl(\frac{\pi d_{ij}}{r_c}\bigr)+1\Bigr)\mathbf{1}[d_{ij}<r_c]\exp\Bigl[-\beta_k\bigl(e^{-d_{ij}}-\nu_k\bigr)^2\Bigr]
\]
with evenly spaced centers $\nu_k$ and per-channel widths $\beta_k$. The initial edge invariant feature is then constructed as:
\[
\mathbf{e}_{ij} = \mathrm{MLP}_d(\boldsymbol{\phi}_{ij}) + \mathrm{MLP}_s(\mathbf{h}_i) + \mathrm{MLP}_t(\mathbf{h}_j)
\]
This tensor $\mathbf{e}_{ij}$ serves as the static edge feature for the subsequent LEFTNet message-passing blocks. Additionally, $\mathbf{e}_{ij}$ is projected via a multi-layer perceptron to generate the multi-head pairwise attention biases required by the Uni-Mol encoder. The equivariant vector features are initialized to zeros, denoted as $\mathbf{v}_i^{(0)} = \mathbf{0}$, and are iteratively updated in the deeper network layers.

The reactant and product initialization modules share learnable parameters to preserve endpoint exchange symmetry. The TS generator uses an independent instantiation of the same initialization design because its input is the dynamically evolving interpolant rather than a fixed reaction endpoint.

\subsection{Learning reaction-transformation-aware representations}
\label{dual_reac_enc}

Starting from the initialized endpoint features, the reaction encoder represents elementary reactions as structural transformations between atom-aligned endpoints. Atom mapping exposes a one-to-one correspondence between reactant and product atoms, allowing the encoder to compare how each local environment changes along the reaction. This comparison is chemically sparse: most atoms and local substructures remain nearly unchanged, while the reaction center undergoes bond rearrangement and geometric distortion. The encoder therefore aims to highlight reaction-relevant structural changes while retaining the conserved scaffold context needed by the generator. Because exchanging the two endpoints reverses the reaction direction without changing the underlying TS geometry, the endpoint branches share weights and communicate through atom-mapped Cross-State Fusion.

To obtain structure-aware invariant features, each branch first processes the initial embeddings $\mathbf{h}_i$ alongside the derived pairwise attention biases through an initial Uni-Mol encoder~\cite{zhou2023unimol}. We denote the resulting enriched invariant features as $\mathbf{s}_i^{(0)}$. Subsequently, these invariant features and the zero-initialized vector features $\mathbf{v}_i^{(0)}$ are fed into a stack of $L$ LEFTNet-style equivariant message-passing blocks~\cite{du2023a}, where geometric updates are performed jointly with the cross-state fusion operations described below. These blocks iteratively update both the scalar ($\mathbf{s}_i^{(\ell)}$) and vector ($\mathbf{v}_i^{(\ell)}$) representations.

To extract this transformation information, the two branches exchange information at initialization and after every equivariant block $\ell$. We term this atom-mapped interaction Cross-State Fusion. The fusion is performed strictly between corresponding atoms (i.e., atoms sharing the same mapping index $i$): for each branch, the endpoint representation being updated is treated as the state representation, while the atom-mapped representation from the counterpart endpoint provides the context representation. The fusion module integrates scalar and vector channels to produce the updated representation, allowing the encoder to distinguish unchanged scaffold regions from atoms whose local environments differ across the reaction. For the scalar channel, we employ symmetric Feature-wise Linear Modulation (FiLM)~\cite{Perez_Strub_deVries_Dumoulin_Courville_2018}:
\[
\tilde{\mathbf{s}}_{R,i}^{(\ell)} = \mathrm{FiLM}_{\text{enc}}^{(\ell)}\bigl(\mathbf{s}_{R,i}^{(\ell)}; \mathbf{s}_{P,i}^{(\ell)}\bigr)
\]
\[
\tilde{\mathbf{s}}_{P,i}^{(\ell)} = \mathrm{FiLM}_{\text{enc}}^{(\ell)}\bigl(\mathbf{s}_{P,i}^{(\ell)}; \mathbf{s}_{R,i}^{(\ell)}\bigr)
\]
where $\mathrm{FiLM}_{\text{enc}}^{(\ell)}(\mathbf{x};\mathbf{c}) = \boldsymbol{\gamma}_{\text{enc}}^{(\ell)}(\mathbf{c})\odot\mathbf{x} + \boldsymbol{\beta}_{\text{enc}}^{(\ell)}(\mathbf{c})$. For the vector channel, equivariant fusion is achieved via a learned encoder-specific linear projection $W_{\text{enc}}^{(\ell)}$ over the concatenated tensors along the channel dimension, preserving $SE(3)$ equivariance:
\[
\tilde{\mathbf{v}}_{R,i}^{(\ell)} = W_{\text{enc}}^{(\ell)}\bigl[\mathbf{v}_{R,i}^{(\ell)} \;\|\; \mathbf{v}_{P,i}^{(\ell)}\bigr]
\]
\[
\tilde{\mathbf{v}}_{P,i}^{(\ell)} = W_{\text{enc}}^{(\ell)}\bigl[\mathbf{v}_{P,i}^{(\ell)} \;\|\; \mathbf{v}_{R,i}^{(\ell)}\bigr]
\]
The same fusion operator is applied in both directions with shared parameters, aligning the representation with the endpoint symmetry of TS generation. In practice, residual connections are applied to these modulated features to ensure stable training. After $L$ blocks, the encoder outputs layer-wise conditioned features $\{\mathbf{s}_R^{(\ell)},\mathbf{v}_R^{(\ell)}\}_{\ell=1}^L$ and $\{\mathbf{s}_P^{(\ell)},\mathbf{v}_P^{(\ell)}\}_{\ell=1}^L$, which serve as multi-scale reaction contexts for the generator.

\subsection{Learning transformation-conditioned transition-state dynamics}
\label{sec:method_generator}

Using the multi-scale reaction contexts from the dual encoder, the generator learns transition-state dynamics rather than a static coordinate correction. It treats the current TS interpolant coordinates $\mathbf{z}_t$ as a generation state and predicts the velocity field that moves this state toward the target geometry. The generator uses the molecular-state initialization described above: an initial Uni-Mol encoder enriches the invariant features, after which an independent stack of LEFTNet-style equivariant message-passing blocks updates the scalar and vector representations. The generator additionally incorporates a learned time embedding $\tau(t)$, which is added to the scalar node features before the invariant and equivariant updates.

At each generator layer, we inject the reactant and product contexts into the generation state through a symmetric reaction-context injection module. The injection is performed according to the atom-mapping index $i$, yielding an updated generation state for subsequent flow-matching updates. By explicitly linking the corresponding atoms across the three states, the TS generator can reuse information from unchanged scaffold regions while allocating modeling capacity to atoms and bonds undergoing structural rearrangement.

The symmetric reaction-context injection module uses a shared three-way fusion operator to respect endpoint exchange symmetry. For the scalar channel, the modulation is defined as:
\[
\tilde{\mathbf{s}}_{t,i}^{(\ell)} = \mathrm{FiLM}_{\text{gen}}^{(\ell)}\bigl(\mathbf{s}_{t,i}^{(\ell)}; \mathbf{s}_{R,i}^{(\ell)}\bigr) + \mathrm{FiLM}_{\text{gen}}^{(\ell)}\bigl(\mathbf{s}_{t,i}^{(\ell)}; \mathbf{s}_{P,i}^{(\ell)}\bigr)
\]
where the shared parameters of $\mathrm{FiLM}_{\text{gen}}^{(\ell)}$ ensure that the operation is symmetric with respect to the reactants and products. The vector channel is similarly fused through an equivariant linear combination:
\[
\tilde{\mathbf{v}}_{t,i}^{(\ell)} = W_{\text{gen}}^{(\ell)}\bigl[\mathbf{v}_{t,i}^{(\ell)} \;\|\; \mathbf{v}_{R,i}^{(\ell)}\bigr] + W_{\text{gen}}^{(\ell)}\bigl[\mathbf{v}_{t,i}^{(\ell)} \;\|\; \mathbf{v}_{P,i}^{(\ell)}\bigr]
\]
As in the encoder, standard residual connections are applied after the fusion steps to facilitate gradient flow. After the final message-passing layer $L$, the updated scalar and vector features, denoted respectively as $\tilde{\mathbf{s}}_{t,i}^{(L)}$ and $\tilde{\mathbf{v}}_{t,i}^{(L)}$, are fed into a gated equivariant readout block~\cite{pmlr-v139-schutt21a}. This block acts as the final velocity predictor, outputting the coordinate update tensor $\Delta\mathbf{z}_t\in\mathbb{R}^{N\times 3}$ that drives the flow-matching ordinary differential equation.

\subsection{Anchor-aligned flow matching in an endpoint-symmetric reaction frame}
\label{sec:method_fm}

The architecture above parameterizes a velocity field for transition-state generation. We train this field with Conditional Flow Matching (CFM)~\cite{lipman2023flow}, defining the probability path in the com-free reaction frame constructed above. This formulation transports a simple base distribution to the target transition state (TS) geometry while keeping training and sampling aligned to the same endpoint-symmetric anchor.

\paragraph*{Training path and objective.}
Let the observable reaction context be defined by the aligned reactant and product in the canonical frame, denoted as $c=(\bar{\bm x}^{R}, \hat{\bm x}^{P})$. The target endpoint for our flow is the aligned transition state, $\bm x_1 = \hat{\bm x}^{TS}$, which resides in the zero-center subspace $\mathcal{X}_0$. 

To initialize the generative trajectory, we draw a sample from a standard isotropic Gaussian prior $\bm \epsilon \sim \mathcal{N}(\bm 0, \bm I)$. To satisfy the com-free constraint, we project this noise into $\mathcal{X}_0$ using the centering operator:
\begin{equation*}
\bar{\bm \epsilon} = \Pi \bm \epsilon = \bm \epsilon - \frac{1}{N}\bm 1 \bm 1^\top \bm \epsilon.
\end{equation*}
In the standard optimal transport formulation, one would align this centered noise directly with the target TS structure via the Kabsch algorithm to minimize the transport cost within the rotational equivalence class of the TS. However, the true TS geometry is unobservable at inference time. Relying on it to align the prior during training would violate causality and create an intractable train-test discrepancy.

To formulate a fully observable and deterministic generation trajectory, we introduce a surrogate alignment mechanism based on the inferable canonical anchor, $\bm x_{\text{anchor}} = \mathcal{F}_{\text{anchor}}(\bar{\bm x}^{R}, \hat{\bm x}^{P})$. We align the com-free noise to this anchor to form our initial flow state:
\begin{equation*}
\bm x_0 = \mathcal{A}(\bar{\bm \epsilon}, \bm x_{\text{anchor}}).
\end{equation*}
Because both the start point $\bm x_0$ and the endpoint $\bm x_1$ are centered and systematically aligned to the same anchor, we construct an exact straight-line interpolation path:
\begin{equation*}
\bm x_t = \psi_t(\bm x_0, \bm x_1) = (1-t)\bm x_0 + t\bm x_1, \qquad t \in [0, 1].
\end{equation*}
This path admits a constant, time-independent conditional vector field:
\begin{equation*}
\bm u_t(\bm x_t \mid \bm x_0, \bm x_1) = \bm x_1 - \bm x_0.
\end{equation*}

Since any affine combination of centered coordinates remains in $\mathcal{X}_0$, the interpolant $\bm x_t$ is strictly com-free. Consequently, the vector field driving the flow must also be com-free. We enforce this explicitly during training by projecting the network's output velocities. The training objective is given by the anchor-aligned CFM loss:
\begin{equation*}
\mathcal{L}_{\mathrm{FM}}(\theta) = \mathbb{E}_{c, \bm \epsilon, t} \left[ \frac{1}{N} \sum_{i=1}^{N} \left\| \Pi \, \bm v_\theta(\bm x_t, t, c)_i - (\bm x_1 - \bm x_0)_i \right\|_2^2 \right],
\end{equation*}
where $\bm v_\theta$ is the predicted velocity field parameterized by the $SE(3)$-equivariant network. This alignment strategy acts as a practical surrogate for optimal transport while maintaining training-inference consistency. A formal mathematical justification demonstrating how this surrogate mechanism approximates optimal transport under equivariance constraints is provided in Supplementary Sec.~\ref{sec:si_anchor_alignment}, and Algorithm~\ref{alg:train_fm} summarizes the corresponding training procedure.

\begin{algorithm}[htbp]
\caption{Training of Anchor-Aligned Com-Free Flow Matching}
\label{alg:train_fm}
\begin{algorithmic}[1]
\Require Dataset of reactions $\mathcal{D} = \{(\bm x^R, \bm x^P, \bm x^{TS})\}$, equivariant network $\bm v_\theta$, total training steps $N_{\text{steps}}$
\For{$\text{step} = 1$ \textbf{to} $N_{\text{steps}}$}
    \State Sample a reaction $(\bm x^R, \bm x^P, \bm x^{TS}) \sim \mathcal{D}$
    \State Center structures: $\bar{\bm x}^S \gets \Pi \bm x^S$ for $S \in \{R, P, TS\}$
    \State Form canonical context: $\hat{\bm x}^P \gets \mathcal{A}(\bar{\bm x}^P, \bar{\bm x}^R)$
    \State Compute anchor: $\bm x_{\text{anchor}} \gets \mathcal{F}_{\text{anchor}}(\bar{\bm x}^R, \hat{\bm x}^P)$
    \State Align target TS: $\bm x_1 \gets \mathcal{A}(\bar{\bm x}^{TS}, \bm x_{\text{anchor}})$
    \State Sample noise $\bm \epsilon \sim \mathcal{N}(\bm 0, \bm I)$ and center: $\bar{\bm \epsilon} \gets \Pi \bm \epsilon$
    \State Align noise to anchor: $\bm x_0 \gets \mathcal{A}(\bar{\bm \epsilon}, \bm x_{\text{anchor}})$
    \State Sample $t \sim \mathrm{Beta}(0.8, 0.8)$
    \State Construct path: $\bm x_t \gets (1-t)\bm x_0 + t\bm x_1$
    \State Construct target field: $\bm u_t \gets \bm x_1 - \bm x_0$
    \State Predict and center velocity: $\hat{\bm u}_t \gets \Pi \, \bm v_\theta(\bm x_t, t, \bar{\bm x}^R, \hat{\bm x}^P)$
    \State Update $\theta$ by minimizing $\frac{1}{N} \sum_{i=1}^N \| \hat{\bm u}_{t, i} - \bm u_{t, i} \|_2^2$
\EndFor
\end{algorithmic}
\end{algorithm}

\paragraph*{Sampling.}
During inference, the model generates the transition state entirely from the observable reactant and product structures. Given a new pair $(\bm x^R, \bm x^P)$, we first center them and align the product with the reactant to form the context $c=(\bar{\bm x}^{R}, \hat{\bm x}^{P})$. We compute the identical canonical anchor $\bm x_{\text{anchor}}$ used during training.

A standard Gaussian sample is drawn and centered to remain in the zero-centered subspace. This com-free noise is then rigidly aligned to the anchor to initialize the sampling trajectory:
\begin{equation*}
\bm x_{t=0} = \mathcal{A}(\Pi \bm \epsilon, \bm x_{\text{anchor}}).
\end{equation*}
The TS geometry is subsequently generated by integrating the learned ordinary differential equation (ODE) using a numerical solver:
\begin{equation*}
\frac{d\bm x_t}{dt} = \Pi \, \bm v_\theta(\bm x_t, t, c), \qquad t \in [0,1].
\end{equation*}
The projection operator $\Pi$ is applied at each solver step to counteract numerical drift, guaranteeing that the trajectory remains strictly within $\mathcal{X}_0$. The terminal state $\bm x_{t=1}$ is returned as the predicted TS conformation. 

The same anchor-aligned initialization is used during training and inference, so the sampling trajectory follows the training construction. The inference procedure is summarized in Algorithm~\ref{alg:sample_fm}.

\begin{algorithm}[htbp]
\caption{Sampling with Anchor-Aligned Com-Free Flow Matching}
\label{alg:sample_fm}
\begin{algorithmic}[1]
\Require Reactant $\bm x^R$, Product $\bm x^P$, network $\bm v_\theta$, ODE solver
\State Center structures: $\bar{\bm x}^R \gets \Pi \bm x^R$, $\bar{\bm x}^P \gets \Pi \bm x^P$
\State Align product: $\hat{\bm x}^P \gets \mathcal{A}(\bar{\bm x}^P, \bar{\bm x}^R)$
\State Compute anchor: $\bm x_{\text{anchor}} \gets \mathcal{F}_{\text{anchor}}(\bar{\bm x}^R, \hat{\bm x}^P)$
\State Sample noise $\bm \epsilon \sim \mathcal{N}(\bm 0, \bm I)$ and center: $\bar{\bm \epsilon} \gets \Pi \bm \epsilon$
\State Initialize path: $\bm x_0 \gets \mathcal{A}(\bar{\bm \epsilon}, \bm x_{\text{anchor}})$
\State Solve ODE: $\frac{d\bm x_t}{dt} = \Pi \, \bm v_\theta(\bm x_t, t, \bar{\bm x}^R, \hat{\bm x}^P)$ from $t=0$ to $t=1$
\State \Return $\bm x_{t=1}$
\end{algorithmic}
\end{algorithm}

\backmatter

\section*{Declarations}

\subsection*{Supplementary information}
Supplementary information is included after the main article in this arXiv version. Two Python files, \texttt{anc/xyz2smiles.py} and \texttt{anc/rmsd.py}, are provided as ancillary files documenting the endpoint-matching and RMSD-evaluation procedures described in the Supplementary Information.

\subsection*{Data availability}

The GDB-10-rxn and GDB-17-rxn out-of-distribution datasets are available via Figshare at \url{https://doi.org/10.6084/m9.figshare.29311196.v2}. The Transition1x dataset is available at \url{https://gitlab.com/matschreiner/Transition1x}, and the RGD1 dataset is available at \url{https://github.com/zhaoqy1996/RGD1}. The processed data used in this work are also available in our GitHub repository at \url{https://github.com/zengkaipeng/TransTS}.

\subsection*{Code availability}
The implementation code is available in the TransTS GitHub repository at \url{https://github.com/zengkaipeng/TransTS}.

\subsection*{Author contribution}
K.Z. proposed the central idea, implemented the code, and wrote the manuscript. J.Z. and S.X. reproduced the baseline methods. W.Z. conducted the quantum-chemical calculations. S.W. performed the mathematical verification and contributed to discussions. B.L. conducted the case study. T.Z. and J.Y. supervised the research.

\clearpage
\section*{Supplementary Information}
\addcontentsline{toc}{section}{Supplementary Information}
\setcounter{section}{0}
\setcounter{subsection}{0}
\setcounter{subsubsection}{0}
\setcounter{figure}{0}
\setcounter{table}{0}
\setcounter{equation}{0}
\renewcommand{\tablename}{Supplementary Table}
\renewcommand{\figurename}{Supplementary Fig.}
\renewcommand{\theHsection}{supp.\arabic{section}}
\renewcommand{\theHsubsection}{supp.\arabic{section}.\arabic{subsection}}
\renewcommand{\theHsubsubsection}{supp.\arabic{section}.\arabic{subsection}.\arabic{subsubsection}}
\renewcommand{\theHfigure}{supp.\arabic{figure}}
\renewcommand{\theHtable}{supp.\arabic{table}}
\renewcommand{\theHequation}{supp.\arabic{equation}}
\section{Data Information}
\label{sec:si_data_information}

\subsection{Construction and Quantum-Chemical Validation of GDB-10-rxn and GDB-17-rxn}
\label{sec:si_ood_dataset_construction}

To rigorously evaluate generalization to unseen chemical space, we constructed two distinct out-of-distribution (OOD) test sets. The first set, GDB-10-rxn, comprises 779 reactions derived from the GDB-10 database~\cite{GDB-13}. This set maintains a similar elemental composition and molecular size to the training data, with molecules containing up to 10 heavy atoms, while introducing reaction types not covered by the training distribution. The second set, GDB-17-rxn, comprises 174 reactions from GDB-17~\cite{GDB-17} and is designed to probe performance on larger and more complex systems containing up to 17 heavy atoms.

The construction workflow for both datasets began by randomly selecting 100 seed molecules from each database. Reaction networks were then enumerated using the YARP package~\cite{YARP2}. For each retained reaction, transition-state geometries were optimized at the $\omega$B97X-D/def2-SVP level, and valid pathways were confirmed by intrinsic reaction coordinate (IRC) calculations. The released dataset also includes IRC-path structures with energy and force labels for reactive machine-learning potential training~\cite{li2025generalReactiveMLP}; however, the present TS-generation benchmark uses only the raw optimized reactant, product, and transition-state geometries required for geometry prediction and evaluation.

\subsection{Post-Checked Transition1x Test Subset}
\label{sec:si_transition1x_post_check}

We constructed a post-checked test subset of Transition1x (T1x) to ensure that the reference TSs used for clean-label analysis correspond to the annotated elementary reactions. Starting from the 287 reactions in the Transition1x test split~\cite{schreiner2022transition1x}, we used each provided reference TS as the initial guess for saddle-point optimization, followed by IRC analysis. Saddle-point optimization and IRC calculations were performed with Gaussian 16 at the $\omega$B97X/6-31G(d) level of theory, matching the original Transition1x calculation level~\cite{gaussian16,wb97x}. Each reference TS was optimized and subjected to frequency analysis using the Gaussian route \texttt{Opt=(TS, CalcFC, ReCalcFC=5, NoEigenTest) Freq}. After saddle-point optimization converged and a single imaginary frequency was confirmed, the final structures were extracted from the Gaussian log files with the Atomic Simulation Environment (ASE)~\cite{ase} and converted into inputs for IRC analysis. IRC calculations used \texttt{IRC=(CalcFC, ReCalc=5, LQA, MaxPoints=80)}, tracing both forward and reverse directions with up to 80 IRC points in each direction.

For each reaction, we optimized the annotated reactant and product geometries and the two IRC endpoints. A reference TS was accepted only if the two IRC endpoints could be assigned to the annotated reactant-product pair. Endpoint identity was determined by converting the optimized XYZ geometries to SMILES using the ancillary file \texttt{anc/xyz2smiles.py}, with atom mapping employed to preserve correspondence between the compared structures. When SMILES conversion failed for either endpoint, the corresponding geometries were matched instead by a 0.1~\AA{} RMSD threshold. Cases that failed during saddle-point optimization or IRC calculations were treated as unmatched.

This reference-TS-initialized post-check retained 262 reactions. For the remaining 25 reactions, we performed an additional recovery check using the ML-predicted TS guesses from all evaluated models as alternative initial guesses and applied the same saddle-point optimization, IRC, and endpoint-matching protocol. This recovered three additional reaction-consistent TSs, yielding a post-checked Transition1x test subset of 265 reactions out of the original 287.

\section{Implementation Details}
\label{sec:si_implementation_details}

\subsection{Training Details}
\label{sec:si_training_details}

Both reported variants were trained with Adam. The learning rate was linearly increased to a peak value of $2\times10^{-4}$ during the warm-up phase and subsequently decayed exponentially with factor $\gamma$. Training lasted for 2000 epochs, with an effective batch size of 256 and validation every 5 epochs. During validation, transition-state samples were generated with the midpoint ODE solver using 10 fixed steps. For checkpoint selection, we minimized the direct validation Kabsch RMSD before atom reindexing. The model-specific settings are summarized in Supplementary Table~\ref{tab:si_training_hparams}.

\begin{table*}[htbp]
\centering
\small
\setlength{\tabcolsep}{5pt}
\caption{Model-specific training settings for the two reported variants. Shared settings are given in the text.}
\label{tab:si_training_hparams}
\begin{tabular}{lcc}
\toprule
\textbf{Hyperparameter} & \textbf{T1x} & \textbf{T1x+RGD1} \\
\midrule
Embedding dimension & 384 & 512 \\
Invariant layers & 2 & 3 \\
SE(3) layers & 5 & 6 \\
Attention heads & 6 & 8 \\
Decay factor $\gamma$ & 0.999 & 0.9993 \\
Warm-up (epochs) & 10 & 20 \\
\bottomrule
\end{tabular}
\end{table*}

\subsection{Inference Details}
\label{sec:si_inference_details}

During inference, the trained flow-matching model generated transition-state initial guesses by integrating the learned ordinary differential equation from the anchor-aligned Gaussian-noise initial state to $t=1$. Numerical integration was performed with \texttt{torchdiffeq}~\cite{torchdiffeq} using the midpoint solver with 15 ODE steps. For a fair comparison, all models, including \modelname{}, were evaluated using one sampled trajectory per reaction.

\subsection{Evaluation Metrics and Transition-State Optimization}
\label{sec:si_evaluation_metrics}

Geometric accuracy was evaluated before saddle-point optimization. Following previous works~\cite{duan2023accurate,duan2025optimal}, RMSD denotes the minimum value after atom reindexing, chirality-insensitive evaluation, and Kabsch rigid alignment~\cite{kabsch1976solution,kabsch1978discussion}. Atom matching used \texttt{pymatgen}~v2026.5.4 and \texttt{pymatgen-core}~v2026.5.18~\cite{ong2013python}. The evaluator implementation is provided as the ancillary file \texttt{anc/rmsd.py}. The genetic atom-order search was unbounded for Transition1x, where molecules are small, and capped at a \texttt{GeneticOrderMatcher} fragment-RMSD threshold of 1.25~\AA{} for the OOD datasets to keep runtime tractable. Direct Kabsch-aligned RMSD was retained as an upper bound. Chirality-insensitive RMSD was obtained by evaluating the mirrored geometry and taking the smaller value. These pre-optimization RMSD values are separate from the downstream quantum-chemical refinement outcomes.

Saddle-point optimizations were carried out with Gaussian 16~\cite{gaussian16}. To match the electronic-structure settings used to generate the corresponding references, Transition1x calculations used the $\omega$B97X/6-31G(d) level of theory~\cite{wb97x}, whereas the OOD test sets used $\omega$B97X-D/def2-SVP~\cite{wb97xd,def2svp}. Each generated TS guess was refined and subjected to frequency analysis with the Gaussian route \texttt{Opt=(TS, CalcFC, NoEigenTest) Freq}. Unless otherwise specified, the maximum number of optimization steps, geometry-convergence criteria, and related optimization settings followed the Gaussian 16 defaults~\cite{gaussian16}. A refinement was treated as a saddle-point success only when the optimized structure had exactly one imaginary frequency.

Post-optimization evaluation was performed on the same generated initial guesses. We report three outcome rates over all reactions in the evaluated set. The invalid rate is the fraction of generated structures that could not be used as quantum-chemical inputs. A generated structure was marked as having a numerical coordinate explosion if, after subtracting the per-axis centroid of the selected prediction, any Cartesian coordinate had an absolute value larger than 50~\AA. Structures for which the \texttt{PySCF}~v2.13.1~\cite{sun2018pyscf} single-point calculation did not reach SCF convergence were also treated as invalid; these convergence checks used the corresponding dataset-level electronic-structure setting stated above. The success rate is the fraction of all reactions that satisfied the saddle-point success criterion above. The correctness rate is the fraction of all reactions for which the optimized saddle point corresponds to the intended input elementary reaction under a 0.1~\AA{} RMSD threshold.

\section{Supplementary Experimental Results}
\label{sec:si_supplementary_experimental_results}

\subsection{Full Benchmark Statistics}
\label{sec:si_full_benchmark_statistics}

To facilitate quantitative comparison across methods, we report the RMSD distribution summaries and post-optimization outcome rates in Supplementary Tab.~\ref{tab:si_rmsd_statistics} and Supplementary Tab.~\ref{tab:si_outcome_rates}. RMSD values are summarized by the median and interquartile range over valid generated structures. Outcome rates are computed over all reactions in each evaluated set.

\begin{table*}[htbp]
\centering
\small
\setlength{\tabcolsep}{4pt}
\caption{Pre-optimization RMSD statistics. Values are median [Q1, Q3] in \AA, where Q1 and Q3 denote the 25th and 75th percentiles. T1x-original uses the original Transition1x labels, whereas T1x-clean uses the post-checked clean subset. Lower median RMSD is better. \textbf{Bold} and \underline{underlined} entries denote the best and second-best results, respectively, within each training setting and metric column.}
\label{tab:si_rmsd_statistics}
\resizebox{\linewidth}{!}{%
\begin{tabular}{c l cccc}
\toprule
\textbf{\makecell{Training\\set}} & \multicolumn{1}{c}{\textbf{Model}} & \textbf{T1x-original} & \textbf{T1x-clean} & \textbf{GDB-10-rxn} & \textbf{GDB-17-rxn} \\
\midrule
\multirow{4}{*}[-0.2em]{\makecell{T1x}} & OA-ReactDiff~\cite{duan2023accurate} & 0.162 [0.061, 0.572] & 0.146 [0.061, 0.561] & 1.095 [0.922, 1.300] & -- \\
 & React-OT~\cite{duan2025optimal} & \textbf{0.100} [0.056, 0.212] & \textbf{0.098} [0.054, 0.207] & \underline{0.401} [0.272, 0.559] & \underline{0.485} [0.362, 0.560] \\
 & MolGEN~\cite{tuo2026flow} & \underline{0.137} [0.069, 0.251] & \underline{0.138} [0.067, 0.254] & 0.445 [0.286, 0.673] & 1.691 [0.725, 1.974] \\
 & \modelname{} & 0.171 [0.097, 0.290] & 0.168 [0.091, 0.291] & \textbf{0.379} [0.270, 0.489] & \textbf{0.429} [0.336, 0.553] \\
\midrule
\multirow{4}{*}[-0.2em]{\makecell{T1x\\+\\RGD1}} & OA-ReactDiff~\cite{duan2023accurate} & 0.092 [0.043, 0.441] & 0.084 [0.043, 0.423] & -- & -- \\
 & React-OT~\cite{duan2025optimal} & 0.083 [0.047, 0.188] & 0.083 [0.046, 0.180] & 0.410 [0.245, 0.627] & 0.582 [0.418, 0.727] \\
 & MolGEN~\cite{tuo2026flow} & \underline{0.074} [0.040, 0.171] & \underline{0.072} [0.040, 0.173] & \underline{0.367} [0.218, 0.645] & \underline{0.564} [0.367, 1.182] \\
 & \modelname{} & \textbf{0.069} [0.036, 0.166] & \textbf{0.067} [0.035, 0.165] & \textbf{0.310} [0.192, 0.457] & \textbf{0.362} [0.262, 0.457] \\
\bottomrule
\end{tabular}
}
\end{table*}

\begin{table*}[htbp]
\centering
\small
\setlength{\tabcolsep}{3pt}
\caption{Post-optimization outcome rates (\%). Invalid, success and correctness rates are computed over all reactions in each evaluated set. Lower invalid rate is better; higher success and correctness rates are better. \textbf{Bold} and \underline{underlined} entries denote the best and second-best results, respectively, within each training setting and metric column.}
\label{tab:si_outcome_rates}
\resizebox{\linewidth}{!}{%
\begin{tabular}{c l ccc c ccc c ccc}
\toprule
\multirow{2}{*}[-0.3em]{\makecell[c]{\textbf{Training}\\\textbf{set}}} &
\multicolumn{1}{c}{\multirow{2}{*}[-0.3em]{\textbf{Model}}} &
\multicolumn{3}{c}{\textbf{T1x}} & &
\multicolumn{3}{c}{\textbf{GDB-10-rxn}} & &
\multicolumn{3}{c}{\textbf{GDB-17-rxn}} \\
\cmidrule(lr){3-5}\cmidrule(lr){7-9}\cmidrule(lr){11-13}
 & & \textbf{Invalid} & \textbf{Success} & \textbf{Correct.} & &
\textbf{Invalid} & \textbf{Success} & \textbf{Correct.} & &
\textbf{Invalid} & \textbf{Success} & \textbf{Correct.} \\
\midrule
\multirow{4}{*}[-0.2em]{\makecell{T1x}} & OA-ReactDiff~\cite{duan2023accurate} & 0.0 & 93.7 & 30.3 & & 64.1 & 27.5 & 0.8 & & -- & -- & -- \\
 & React-OT~\cite{duan2025optimal} & \underline{0.0} & \textbf{96.5} & \textbf{73.2} & & 0.3 & \textbf{82.9} & \underline{41.8} & & 6.3 & \underline{59.2} & \underline{24.7} \\
 & MolGEN~\cite{tuo2026flow} & 0.7 & 92.3 & 39.4 & & \underline{0.1} & 72.9 & 21.4 & & \underline{1.1} & 20.1 & 3.4 \\
 & \modelname{} & \textbf{0.0} & \underline{94.1} & \underline{71.8} & & \textbf{0.1} & \underline{80.6} & \textbf{55.1} & & \textbf{0.6} & \textbf{63.2} & \textbf{37.9} \\
\midrule
\multirow{4}{*}[-0.2em]{\makecell{T1x\\+\\RGD1}} & OA-ReactDiff~\cite{duan2023accurate} & 0.0 & \underline{97.2} & 34.8 & & 100.0 & 0.0 & 0.0 & & -- & -- & -- \\
 & React-OT~\cite{duan2025optimal} & 0.0 & \textbf{97.6} & \underline{72.5} & & 0.3 & \underline{86.0} & \underline{48.8} & & \underline{0.6} & \underline{56.9} & \underline{21.3} \\
 & MolGEN~\cite{tuo2026flow} & \underline{0.0} & 94.4 & 41.1 & & \underline{0.3} & 77.0 & 22.3 & & 1.1 & 52.3 & 16.1 \\
 & \modelname{} & \textbf{0.0} & 96.5 & \textbf{75.3} & & \textbf{0.1} & \textbf{87.8} & \textbf{61.5} & & \textbf{0.0} & \textbf{83.3} & \textbf{58.6} \\
\bottomrule
\end{tabular}
}
\end{table*}

\subsection{OOD Bond-Rearrangement Stratification}
\label{sec:si_ood_bnfm_stratification}

For the OOD bond-rearrangement analysis, reactions were assigned to graph-defined B$n$F$m$ classes, where $n$ is the number of broken bonds and $m$ is the number of formed bonds between the atom-mapped reactant and product graphs. Reactant and product graphs were obtained from the optimized endpoint geometries using the same ancillary file \texttt{anc/xyz2smiles.py} used for the IRC endpoint-matching procedure in Supplementary Sec.~\ref{sec:si_transition1x_post_check}. All mapped atoms and bonds, including hydrogen-involving bonds, were included. A removed bond contributed one broken bond, a newly formed bond contributed one formed bond, and a bond-order change contributed one broken and one formed bond. The B0F0 class therefore denotes cases for which this graph-level comparison detects no broken, formed, or bond-order-changed bonds, while the endpoints can still differ through conformational or stereochemical rearrangements, including chirality-related changes, planar isomerism, or transitions between low-energy conformers. Reactions for which valid reactant/product SMILES could not be generated were assigned to a mandatory standalone \texttt{NO\_SMILES} class.

To keep the stratified comparison interpretable while retaining the dominant bond-rearrangement classes, long-tail classes were compressed independently for GDB-10-rxn and GDB-17-rxn. The \texttt{NO\_SMILES} class was always retained as a standalone class. The remaining chemical B$n$F$m$ classes were sorted by decreasing reaction count and retained until the cumulative coverage, including \texttt{NO\_SMILES}, reached at least 85\% of the dataset. If the cutoff count was tied, all classes with that count were retained. The remaining chemical classes were merged into \texttt{Other}, with \texttt{NO\_SMILES} excluded from this merge. We additionally required the merged \texttt{Other} group to be smaller than the largest retained exact B$n$F$m$ class.

The resulting class distributions are shown in Supplementary Fig.~\ref{fig:si_bnfm_reaction_counts}. The retained classes cover 86.78\% of GDB-10-rxn and 87.36\% of GDB-17-rxn, with the remaining long-tail chemical classes collected into \texttt{Other}. Because several exact B$n$F$m$ classes contain only a small number of reactions, the stratified post-optimization comparisons are reported as absolute counts rather than within-class rates. The corresponding stratified optimization-success counts, reaction-correctness counts, and median pre-optimization RMSD values are reported in Supplementary Figs.~\ref{fig:si_bnfm_success_counts}--\ref{fig:si_bnfm_median_rmsd}.

\begin{figure*}[htbp]
    \centering
    \includegraphics[width=\textwidth]{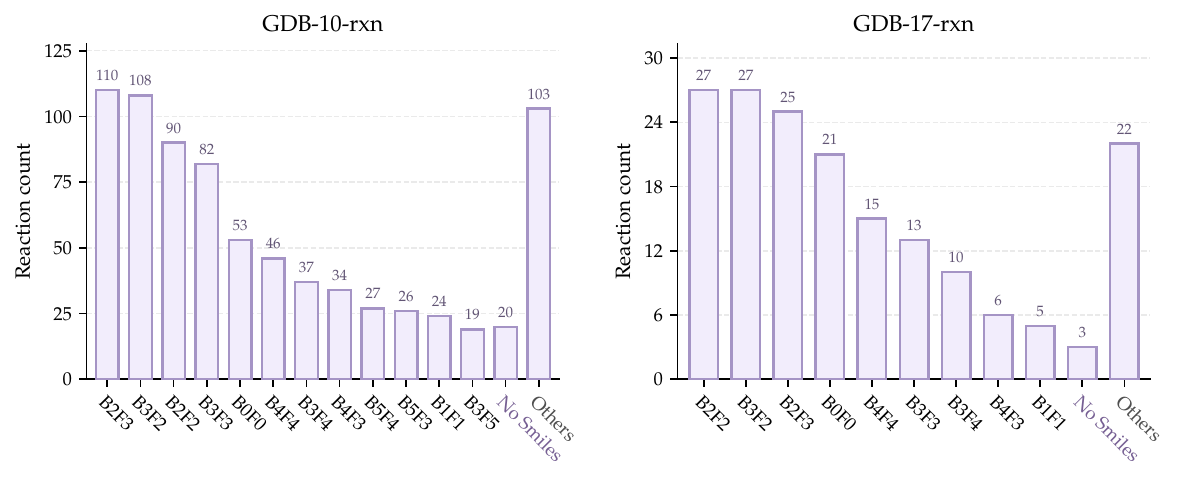}
    \caption{\textbf{OOD bond-rearrangement class distributions.}
    Reaction counts are shown for the retained graph-defined B$n$F$m$ classes in GDB-10-rxn and GDB-17-rxn. Percentages use the full dataset sizes as denominators: 779 reactions for GDB-10-rxn and 174 reactions for GDB-17-rxn. The \texttt{NO\_SMILES} class is retained as a standalone class, whereas \texttt{Other} contains the compressed long-tail chemical B$n$F$m$ classes.}
    \label{fig:si_bnfm_reaction_counts}
\end{figure*}

\begin{figure*}[htbp]
    \centering
    \includegraphics[width=\textwidth]{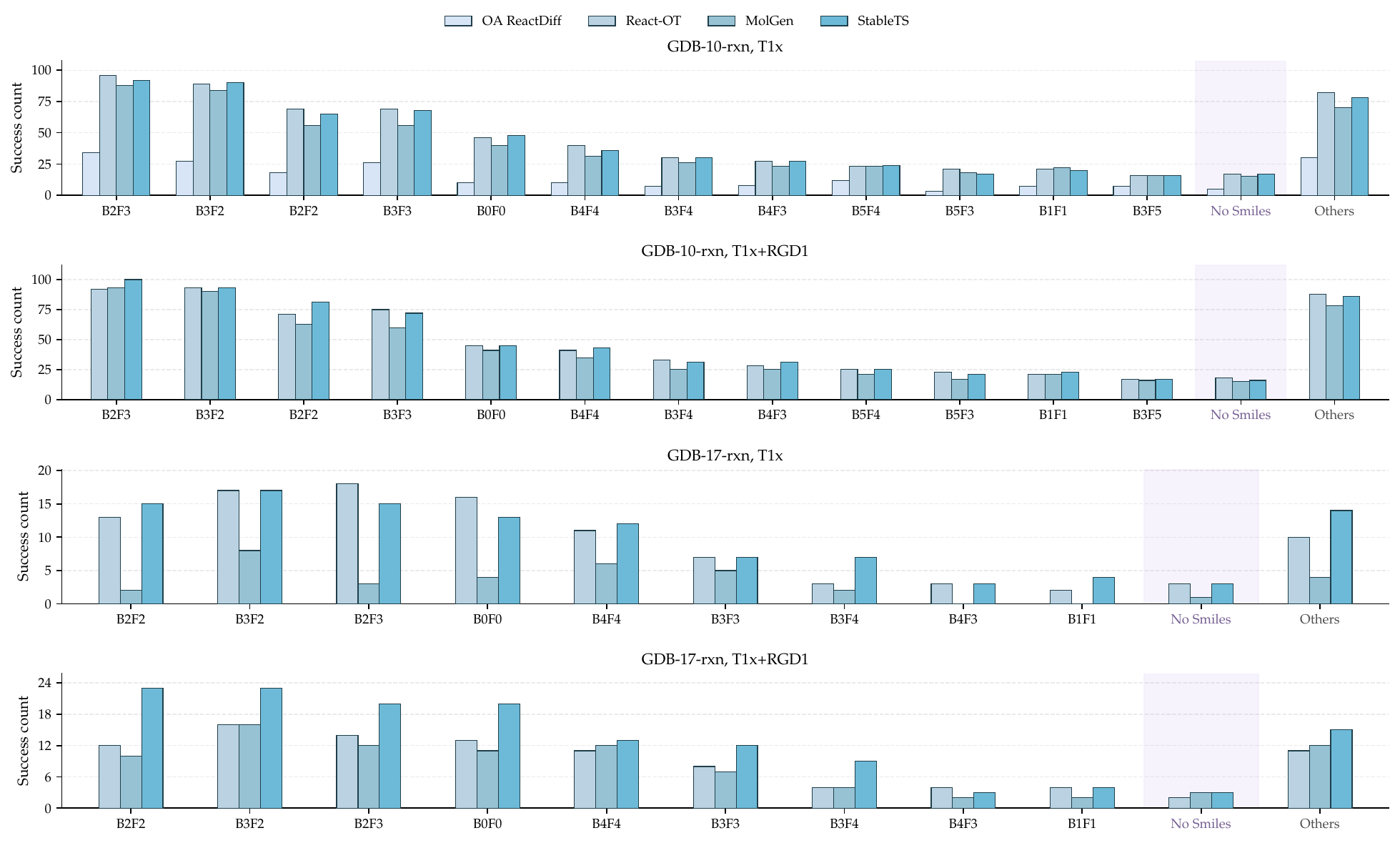}
    \caption{\textbf{Optimization-success counts stratified by OOD bond-rearrangement class.}
    Absolute success counts are shown after saddle-point optimization. Each row corresponds to one OOD dataset and training-set combination, and bars compare \modelname{} with baseline TS-generation models within each retained B$n$F$m$ class. OA-ReactDiff is not shown for GDB-10-rxn after T1x+RGD1 training or for either GDB-17-rxn training setting because it produced no valid structures in these settings.}
    \label{fig:si_bnfm_success_counts}
\end{figure*}

\begin{figure*}[htbp]
    \centering
    \includegraphics[width=\textwidth]{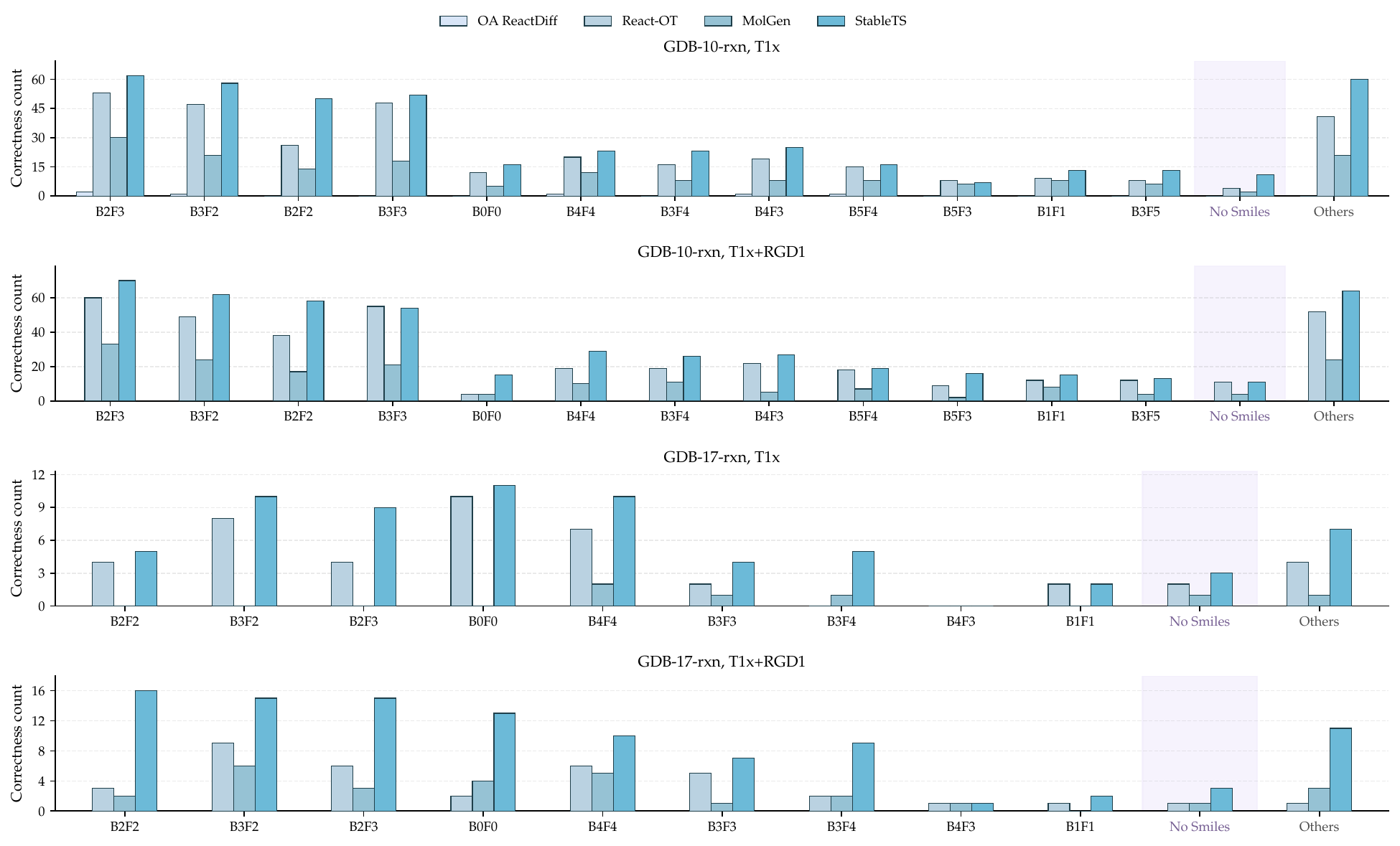}
    \caption{\textbf{Reaction-correctness counts stratified by OOD bond-rearrangement class.}
    Absolute correctness counts are shown after saddle-point optimization. Each row corresponds to one OOD dataset and training-set combination, and bars compare \modelname{} with baseline TS-generation models within each retained B$n$F$m$ class. OA-ReactDiff is not shown for GDB-10-rxn after T1x+RGD1 training or for either GDB-17-rxn training setting because it produced no valid structures in these settings.}
    \label{fig:si_bnfm_correctness_counts}
\end{figure*}

\begin{figure*}[htbp]
    \centering
    \includegraphics[width=\textwidth]{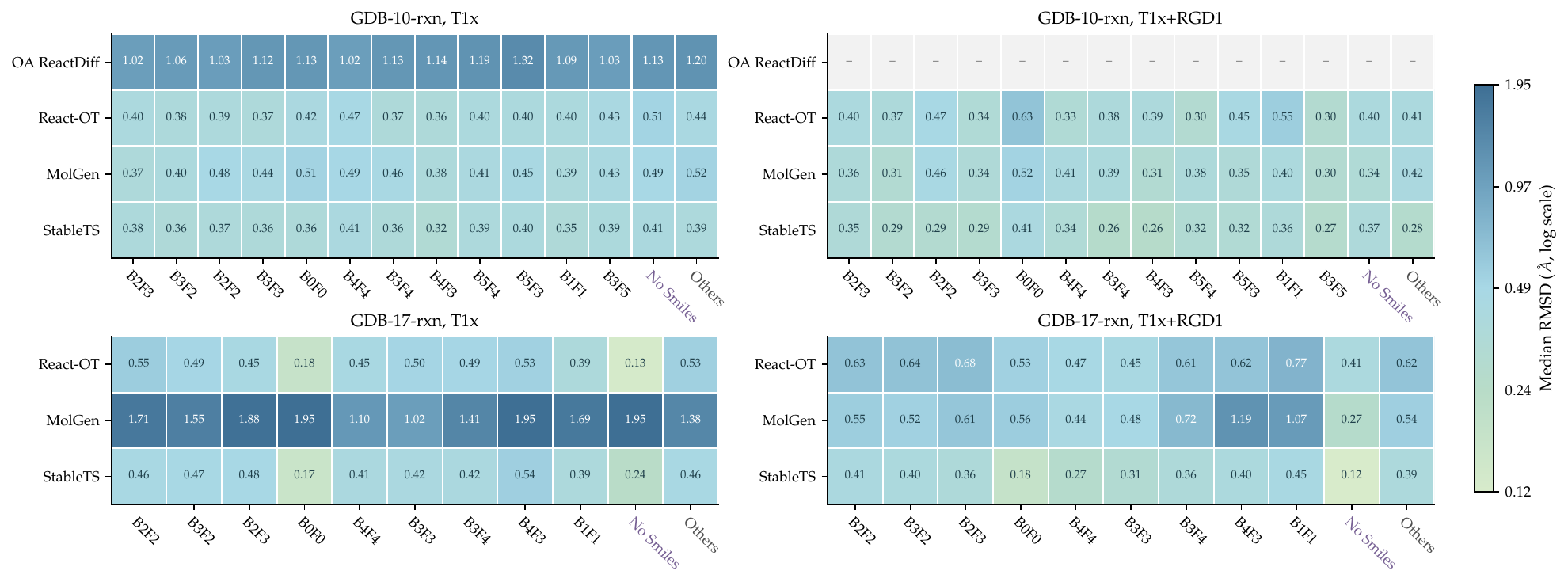}
    \caption{\textbf{Median pre-optimization RMSD stratified by OOD bond-rearrangement class.}
    Heatmaps report the median pre-optimization RMSD for each model and retained B$n$F$m$ class. Rows of panels correspond to GDB-10-rxn and GDB-17-rxn, and columns correspond to T1x and T1x+RGD1 training. Lower RMSD values indicate closer coordinate-level agreement with the reference TS before saddle-point optimization. Blank entries indicate settings with no valid RMSD values after filtering or settings in which the model produced no valid structures.}
    \label{fig:si_bnfm_median_rmsd}
\end{figure*}

\section{Justification for Anchor Alignment}
\label{sec:si_anchor_alignment}

A fundamental challenge in applying flow matching to 3D molecular structures is handling global $SE(3)$ symmetries, particularly when defining optimal transport paths. This section formally justifies our use of anchor-based alignment as a causality-preserving surrogate for optimal transport that strictly maintains train-test consistency.

\paragraph*{The Optimal Transport Ideal.}
In the context of generative modeling for 3D structures, constructing a well-conditioned probability path requires properly addressing the rotational invariance of the target domain. Let $\bm{X}_0 \in \mathbb{R}^{N \times 3}$ be a sample drawn from the base distribution $p_0$ (e.g., zero-mean Gaussian noise), and let $\bm{X}_{\text{ref}} \in \mathbb{R}^{N \times 3}$ denote the ground-truth reference structure. The target distribution $p_1$ is defined on the $SO(3)$ orbit of $\bm{X}_{\text{ref}}$, meaning any valid target state can be expressed as $\bm{X}_1 = \bm{X}_{\text{ref}}R$ for $R \in SO(3)$.

To minimize the dynamical transport cost, we seek an optimal coupling between the noise sample $\bm{X}_0$ and the target orbit under the squared Frobenius norm. For a given $\bm{X}_0$, the optimal transport (OT) plan reduces to finding the optimal rotation $R^*$ that minimizes the spatial displacement:
\begin{equation*}
    R^* = \arg\min_{R \in SO(3)} \| \bm{X}_0 - \bm{X}_{\text{ref}}R \|_F^2
\end{equation*}
Expanding the squared Frobenius norm yields $\|\bm{X}_0\|_F^2 + \|\bm{X}_{\text{ref}}\|_F^2 - 2\text{Tr}(\bm{X}_0^T \bm{X}_{\text{ref}} R)$. Because the norms $\|\bm{X}_0\|_F^2$ and $\|\bm{X}_{\text{ref}}R\|_F^2$ are constant with respect to $R$, minimizing the transport cost is strictly equivalent to maximizing the cross-covariance trace, which is the classic Orthogonal Procrustes problem:
\begin{equation*}
    R^* = \arg\max_{R \in SO(3)} \text{Tr}(\bm{X}_0^T \bm{X}_{\text{ref}} R)
\end{equation*}
This admits an efficient, closed-form solution via Singular Value Decomposition (SVD). By computing the covariance matrix $C = \bm{X}_{\text{ref}}^T \bm{X}_0 = U \Sigma V^T$, the optimal rotation is analytically given by $R^* = V U^T$ (with a determinant correction applied to ensure $R^* \in SO(3)$).

Once the target $\bm{X}_1^* = \bm{X}_{\text{ref}} R^*$ is perfectly aligned with the initial noise $\bm{X}_0$, the $L^2$-Wasserstein optimal transport geometry dictates a displacement interpolation (i.e., a Euclidean straight line) between the coupled pairs:
\begin{equation*}
    \bm{X}_t = (1-t)\bm{X}_0 + t \bm{X}_1^*
\end{equation*}
Consequently, the conditional vector field driving this transformation simplifies to a time-independent, constant velocity:
\begin{equation*}
    v_t(\bm{X}_t \mid \bm{X}_0, \bm{X}_1^*) = \bm{X}_1^* - \bm{X}_0
\end{equation*}
By regressing against this OT-aligned, constant-velocity target, the neural network is relieved of the need to learn complex macroscopic rotational trajectories. This minimal-curvature path dramatically reduces the complexity of the learned vector field and mitigates structural collapse during the intermediate steps of the ODE integration.

\paragraph*{The Observability Constraint.}
While optimal for minimizing trajectory curvature, direct alignment to $\bm{X}^{TS}$ introduces a fundamental causality issue: the ground-truth TS structure $\bm{X}^{TS}$ is unobservable at inference. Relying on it to define the boundary condition $\bm{x}_0 = \mathcal{A}(\bm{\epsilon}, \bm{X}^{TS})$ during training establishes an unresolvable dependence on the target. If a model is trained using this target-aware alignment, it will encounter a severe distribution shift when presented with unaligned noise at inference, as the true optimal rotation $R^*$ cannot be computed without the target.

\paragraph*{Equivariance and the Jointly Moving System.}
To resolve this without sacrificing the efficiency of aligned transport, we leverage the whole-reaction $SE(3)$ equivariance of our architecture. We formulate the reactant, product, and TS as a \textit{jointly moving system}. For any rotation $Q \in SO(3)$, our model $f_\theta$ obeys:
\begin{equation*}
f_\theta(Q \bm{x}^R, Q \bm{x}^P, Q \bm{\epsilon}) = Q f_\theta(\bm{x}^R, \bm{x}^P, \bm{\epsilon}).
\end{equation*}
This property implies that the exact global orientation of the reaction system is mathematically arbitrary, provided the relative geometries among the inputs and the target are preserved. Therefore, any canonical reference frame can be used, as long as it is deterministically constructible exclusively from the observable inputs $(\bm{x}^R, \bm{x}^P)$.

\paragraph*{The Surrogate Anchor Alignment.}
We construct an inferable anchor $\bm{x}_{\text{anchor}} = \mathcal{F}_{\text{anchor}}(\bm{x}^R, \bm{x}^P)$. Because reversing the reactant and product endpoints describes the same transition-state geometry, the anchor is required to be endpoint-symmetric:
\begin{equation*}
\mathcal{F}_{\text{anchor}}(\bm{x}^R, \bm{x}^P)
=
\mathcal{F}_{\text{anchor}}(\bm{x}^P, \bm{x}^R).
\end{equation*}
Thus, the surrogate canonical frame is aligned with the intrinsic endpoint symmetry of transition-state generation rather than being tied to an arbitrary forward or reverse reaction direction. By aligning both the target $\bm{X}^{TS}$ and the sampled noise $\bm{\epsilon}$ to this anchor:
\begin{align*}
\bm{x}_1 &= \mathcal{A}(\bm{X}^{TS}, \bm{x}_{\text{anchor}}), \\
\bm{x}_0 &= \mathcal{A}(\bm{\epsilon}, \bm{x}_{\text{anchor}}),
\end{align*}
we achieve two critical properties:
\begin{enumerate}
    \item \textbf{Causality and Consistency:} $\bm{x}_0$ is computed using only the noise $\bm{\epsilon}$ and the observable anchor. Thus, the exact same initialization procedure can be executed identically during both training and inference, preserving strict causality.
    \item \textbf{Surrogate Optimal Transport:} By forcing $\bm{x}_0$ and $\bm{x}_1$ to inhabit the same canonical frame defined by the reaction endpoints, we drastically reduce the rotational variance of the transport path. While it serves as a surrogate for the true optimal transport (which would require inaccessible target alignment), it effectively minimizes the transport cost within the relative frame of the chemical reaction, resulting in simpler vector fields and significantly faster ODE integration times.
\end{enumerate}
In summary, aligning with the structural anchor utilizes the equivariant properties of the jointly moving system to lock the relative poses of the noise and the target, offering a highly efficient transport trajectory while maintaining strict train-test consistency.

\bibliography{sn-bibliography}
\end{document}